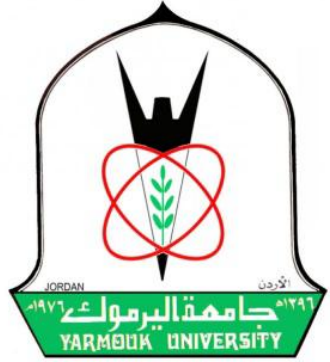

# Yarmouk University

Hijjawi Faculty for Engineering Technology

Communications Engineering Department

"Adaptive Quorum-Based Channel-Hopping Distributed Coordination Scheme for Cognitive Radio Networks"

Prepared by:

Esraa Zeyad Al Jarrah

| Advisor | Co. Advisor |
|---|---|
| Dr. Haythem Bany salameh | Dr. Ali Eyadeh |

2015/2016

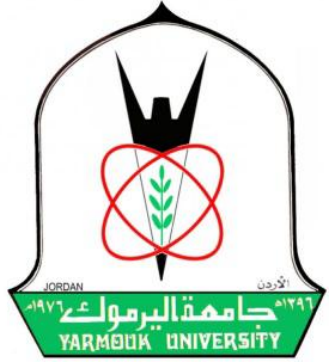

# "Adaptive Quorum-Based Channel-Hopping Distributed Coordination Scheme for Cognitive Radio Networks"

## Yarmouk University

## Hijjawi Faculty for Engineering Technology

March, 2016

# Adaptive Quorum-Based Channel-Hopping Distributed Coordination Scheme for Cognitive Radio Networks

By

Esraa Zeyad Al Jarrah

Thesis Submitted in Partial Fulfillment of the Requirements for the Degree of Master of Science in Electrical Engineering

At

Hijjawi Faculty for Engineering Technology

Yarmouk University

March, 2016

| Committee Members | Signature and Date |
|---|---|
| Dr. Haytham Bany Salameh (*Advisor*) | 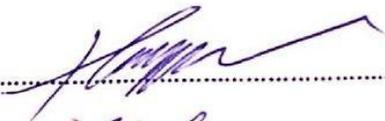 |
| Dr. Ali Eyadeh (*Co.Advisor*) | 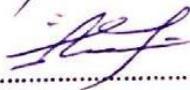 |
| Dr. Ahmad Al-Mousa (*Member*) | 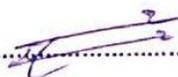 |
| Dr. Taimour Al-dalgamouni (*Member*) | 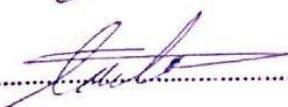 |

# Dedication

*This thesis is dedicated to my great parents who have always loved me unconditionally and whose good examples have taught me to work hard for the things that I aspire to achieve. To Mohammad, my guardian angel and heavenly beautiful gift. To my beloved brothers and sisters, the symbol of love and giving. To my friends who encourage and support me. To all the people in my life who touch my heart.*

# Acknowledgment

"... يَرْفَعِ اللَّهُ الَّذِينَ آمَنُوا مِنكُمْ وَالَّذِينَ أُوتُوا الْعِلْمَ دَرَجَاتٍ وَاللَّهُ بِمَا تَعْمَلُونَ خَبِيرٌ " سورة المجادله - الآيه 11.

*First and foremost, I thank Allah, whose many blessings have made me to do this work and for giving me the ability to complete it successfully.*

*I would like to express my deepest gratitude to my advisor Dr. Hythem Bany Salameh for his excellent guidance, patience and immense knowledge. His guidance helped me in all the time of research and writing of this thesis.*

*I am also would like to thank Dr. Ali Eyadeh who accepted to be my Co.Adviser; I want to thank him for his support for getting this work done.*

*My thanks and gratitude also go to my committee members Dr. Ahmad Al-Mousa & Dr. Taimour Al-dalgamouni, and also to the academic committee of the telecommunication department at the Hijjawi factualty for engineering technology.*

*All Praise is to Allah*
*Esraa Zeyad Al-Jarrah*

# Table of Contents

# List of Figures:

# List of Abbreviations:

| | |
|---|---|
| AMRCC | Adaptive Multiple Rendezvous Control Channel. |
| CR | Cognitive Radio. |
| CRN | Cognitive Radio Network. |
| CCC | Common Control Channel. |
| CH | Channel Hopping. |
| CQM | Cyclic-Quorum based multi-channel MAC. |
| CH | Cluster-Head. |
| CTS | Clear to Send. |
| Diag | Diagonal. |
| DTS | Decide to Send. |
| ETTR | Expected Time to Rendezvous. |
| MAC | Medium Access Control. |
| MTTR | Maximum TTR. |
| PUs | Primary Users. |
| POP | Pair-on-Pair. |
| PS | Power-Saving. |
| PR | Primary Radio |
| QRCH | Quorum Rendezvous Channel-Hopping. |
| QCH | Quorum-based Channel-Hopping. |
| QS | Quorum System. |
| RDV | Rendezvous. |
| RCP | Rotation Closure Property. |
| RTS | Request to Send. |
| SUs | Secondary Users. |
| TTR | Time to Rendezvous. |
| UWB | Ultra-Wideband. |

# Abstract

**Adaptive Quorum-Based Channel-Hopping Distributed Coordination Scheme for Cognitive Radio Networks.**
**By Esraa Zeyad Al Jarrah (2013976006).**
**Advisor                                              Co. Advisor**
**Dr. Haythem Bany salameh.             Dr. Ali Eyadeh.**

One of the most important challenges in deploying cognitive radio networks (CRNs) is to find a common control channel (CCC) to all secondary users (SUs) that enables efficient CR communications. This challenge is attributed to the dynamic time-varying change of network topology, location and spectrum availability conditions. Rendezvous, which is the process of establishing control communications, is an essential requirement to enable efficient communication between any two pair of CR nodes. The most popular CR rendezvous protocols are based on quorum systems (QSs). Quorum systems are systematic approaches, which have several attractive properties that can be utilized to establish communication without the need of a CCC and so overcome the rendezvous (RDV) problem.

In this thesis, we propose new channel-hopping-based distributed rendezvous algorithm based on grid-based-quorum techniques. The proposed algorithm increases the probability of RDV within a single cycle by allowing CR nodes to meet more often according to intersection property of quorum systems. Our proposed algorithm is called Adaptive_Quorum-Based Channel-Hopping Distributed Coordination Scheme for Cognitive Radio Networks. The main idea of our algorithm is to dynamically adjusting the selected QS by CR users according to the varying traffic loads in the CRN. The proposed algorithm decreases the average time to rendezvous (TTR) and increase the probability of RDV. We evaluate the performance of our algorithm through Matlab simulations. The performance of proposed algorithm is compared with two different design scheme. The results show that our algorithm can reduce TTR, increase the RDV, and decrease the energy consumption per successful RDV.

# Chapter 1: Introduction:

## 1.1 Overview

In recent years, the increasing demand on the various wireless services resulted in overcrowding the unlicensed radio spectrum. In contrast, measurement studies indicated that the allocated licensed spectrum is vastly under-utilized [1]. To deal with the spectrum scarcity problem, there is a need for a new communication technology that opportunistically exploits the unoccupied wireless spectrum. This technology is known as cognitive radio (CR), which has been developed as an enabling technology for dynamic opportunistic spectrum access. In addition, CRs can provide mobile users with high bandwidth through heterogeneous wireless architectures.

In CR networks (CRN), the unused spectrum by primary users (PUs) is known as spectrum holes. Unlicensed users, which are also called secondary users (SUs), can exploit those spectrum holes to significantly improve the overall spectrum utilization.

The main needed functions for CR are: (1) spectrum sensing, where the SUs can determine the unused spectrum, and (2) spectrum management which deals with finding the best available channels for communication, (3) spectrum mobility, where the SUs must vacate the chosen channel whenever a PU signal is detected and (4) spectrum sharing to provide fair access to the available channels among coexisting SUs. According to the opportunistic channel access mechanism, SUs must dynamically sense the spectrum holes. If a spectrum hole is not occupied by a PU, it can be used by SUs to establish communication links [1].

A CR is a software-defined radio, in which each node can change its operating parameters according to the surrounding RF environment. There are two main characteristics that can be defined from this definition for CR. The first one, is the cognitive capability that enables the radio technology to sense the information from the surrounding environment to determine suitable communication parameters and dynamically adapt it to radio environment. In open spectrum there are several tasks required for the adaptive process, which is called cognitive cycle as shown in Figure 1.

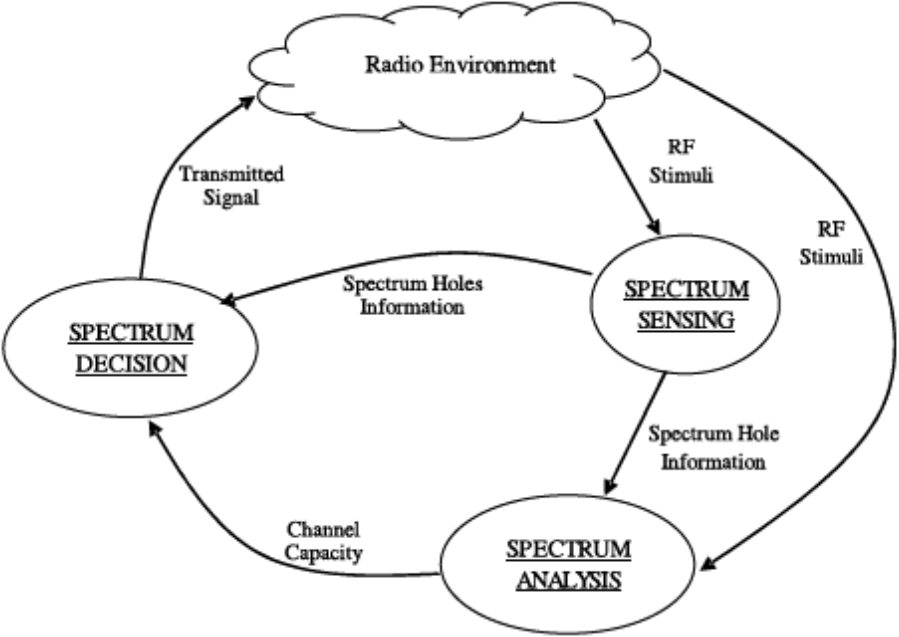

Fig.1: Cognitive cycle [1].

The cognitive cycle contains three main steps [1, 2]:

1. Spectrum sensing: is one of the important requirements in CR, where the available spectrum bands are monitored to detect spectrum holes.

2. Spectrum analysis: the different characteristics of the detected available spectrum holes are analyzed.
3. Spectrum decision: based on the SUs requirement, a CR can determine the transmission bandwidth, the data rate and the transmission mode. Then, the suitable frequency bands are chosen according to the user requirements and decision rule.

The second character, is the CR reconfigurability, which is the ability to adjust the parameters during the radio operation according to the radio environment. There are various reconfigurable parameters that can be adjusted by the CRs (e.g., operating frequency, modulation and transmission power). CR transmission parameters can be reconfigured at the beginning of a transmission, and/or during the transmission.

When two SUs wish to communicate with each other in a CRN, they must first meet on an unoccupied common channel to exchange control messages. This process is called "rendezvous". The rendezvous process is an essential part in CRN operation. Before starting the rendezvous process, users are even not aware of the existence of each other or the number of available channels which are dynamically changing. This makes the implementation of the rendezvous process a challenging problem.

The common control channel (CCC) is used to facilitate the transmission coordination, neighbor discovery, routing information updates and topology changes. CRNs, users maintain networks connectivity by using CCC to broadcast control message for neighboring users. CCC can improve PUs detection by supporting spectrum sensing coordination, which improves spectrum efficiency and CR throughput.

Most of existing studies rely on the existence of a dedicated CCC, which facilities the rendezvous process [3-5]. However, the presences of a dedicated CCC suffers from two main problems. The first one is the bottleneck problem when there is a large number of packets that the SUs want to transmit through the CCC. This can cause a high packet collisions and reduce the efficiency of channel utilization. In addition, it is very difficult to maintain a single dedicated CCC due to the dynamic nature of PUs activities and spectrum heterogeneity. CCC is affected by PUs activity since at any given time, PUs can suddenly appear. Once PUs return to the CCC, it will be difficult to negotiate establishing a new CCC as the users are unable to use the original one. Thus, it is very difficult to find a common available channel to all users because of spectrum heterogeneity.

There is an alternative method to implement the rendezvous without the need of a dedicated CCC, which is using channel hopping protocols [6]. Recently, several quorum-based channel hopping schemes (e.g. [6, 7]) have been proposed to overcome the rendezvous (RDV) problem. The concept of quorum has been widely used for a consistent data replication and agreement problems [8, 9].

Note that power-saving protocols are considered as one of the main applications that use the quorum-based schemes [10-12].

## 1.2 Literature Survey:

The best way to understand the CCC problem is to study the CCC classification. Figure 2 shows the CCC design classification in CRNs [13]. The CCC design schemes are divided into overlay and underlay approaches, (they are distinguished by the way that SUs share the spectrum with PUs). For overlay approaches, an unused PU spectrum channel is allocated for CCC. Once the PU returns to the allocated CCC, the SU must vacate the current channel and reestablish a new CCC.

The overlay approaches are categorized as in-band and out-of-band schemes [2]. For the in-band schemes, the coverage of CCC is locally limited, where out-of-band CCC has a global coverage and can be always available. In band approach is divided to sequence-based and group based schemes. Out-of-band approaches are primarily based on dedicated CCCs. For underlay approaches, the PUs and CCC can be in the same spectrum band. Spread spectrum techniques are utilized in this approach by using short pulses to transmit control messages in low power. The short pulses are spread over a huge bandwidth, which seems as a noise to PUs. The ultra-wideband (UWB) transmission technology is mainly utilized in the underlay CCC approaches.

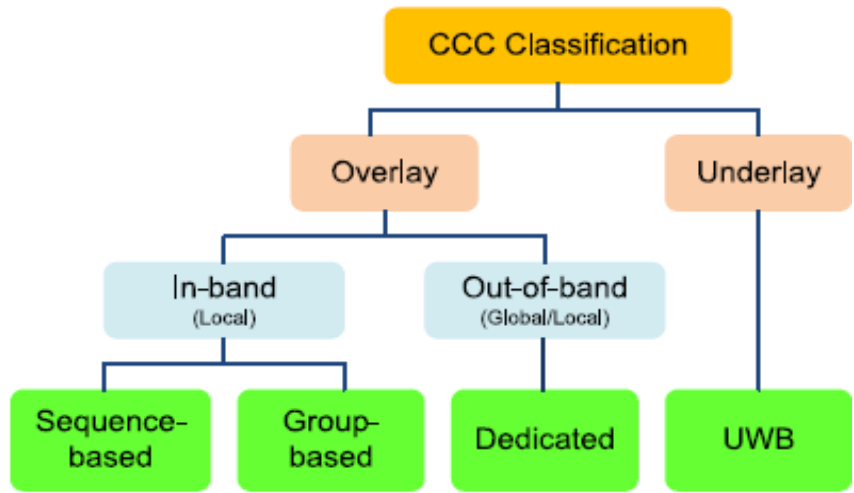

Fig.2: Common control channel scheme classification [13].

Four major control channel design schemes are recognized based on CCC classification: sequence-based [6, 7], [14-22], and [25], group-based [23] and [24], dedicated CCC [41], and underlay CCC [43].

In sequence-based CCC design, the allocation of CCC can be according to predetermined or random channel hopping sequence. According to this design, the impact of PU activity is minimized. Hopping sequence construction can be blind or pseudo-random [14, 15], permutation-based [18], adaptive frequency hopping [25], or quorum-based [6, 7], [19-22].

Blind or pseudo random rendezvous processes [14-17] are considered under the channel hopping design. In [16], the authors proposed a jump-stay based channel-hopping algorithm. This approach is based on generating channel hopping (CH) sequence in rounds, and each of these rounds consists of two patterns (one called jump-pattern and the other is called stay-pattern). Initially, the jump-pattern is performed by each user. In this pattern, users continuously "jump" on the available channels. Then, they follow a stay-pattern, where each user "stays" on a specific channel. The ring-walk based rendezvous algorithm was proposed in [17]. According to this scheme, each channel is represented as a vertex in a ring. In this case, the CH sequence is generated by visiting vertices in the ring. Users walk on the ring in a clockwise direction by visiting the vertices (channels in CRNs) with different velocities. The rendezvous is guaranteed when users with lower velocities will meet with the users with higher velocities. In [17], the authors assumed that each user has a unique ID for designing its velocity and determines the upper bound on network size. We note that in the two [10] and [17] approaches, as the number of channels increases, the TTR increases under asymmetric channel model.

In permutation sequence-based scheme [18], SUs using the permutation of available channels to construct the channel hopping sequences and then increase the probability of RDV between two users. The authors showed that the expected TTR (ETTR) has an upper bound. When the CRN contains a large number of available channels, the neighboring users may take very long time to find each other and to exchange control messages. Once PU returns to a channel, this channel will be removed from the hopping sequences. On the other hand, the hopping sequence in [18] is not flexible to accommodate new channel opportunities (because it is predefined). Also, the authors did not study the case when two SUs observe different available channels. As a result, the two SUs may take longer time to rendezvous, which makes the ETTR not bounded.

In [25], the authors proposed a frequency hopping scheme, which is called an adaptive multiple rendezvous control channel (AMRCC). The proposed hopping sequence scheme is adaptive based on periodic sensing results, where the channels with less introduced interference to PUs will appear more often on the sequence. Also, in this scheme channels of higher quality are allocated for longer slots so that the probability of rendezvous will be increased. The drawback of this work is when a large number of available channels appear in sequence, the CCC link establishment time will be very long (i.e., the TTR may not be bounded).

Quorum-based schemes can be divided into two groups; the first group is so-called quorum system-based protocols, which were proposed for multi-Channel Medium Access Control (MAC) networks [19]. These protocols are based on cyclic quorum systems. In [19], the authors proposed a cyclic-quorum based multi-channel MAC protocol (called CQM). This protocol uses only one transceiver to achieve multiple-rendezvous. In this protocol, each user can switch among different channels based on a prespecified channel hopping scheme. If two users switch to the same channel simultaneously, they can communicate with each other. The CQM protocol can guarantees that any sender can communicate with its intended receiver in a timely manner, which solves the missing receiver problem. Quorum Rendezvous channel-hopping (QRCH) scheme uses quorum for rendezvous in a jamming environment. A novel quorum-channel mapping strategy is exploited by QRCH. Specifically, the elements in a selected quorum are mapped into channel indices. Each node randomly selects a quorum, which its elements are mapped into channel index and then generate two different channel hopping sequences (i.e., a sending sequence and a receiving sequence). Using a quorum system, users are independently able to hop over random channels [20].

The second group belongs to the quorum-based protocols, which is proposed for CRNs [6, 7] and [21]. The quorum-based channel-hopping (QCH) system in [6] introduces three different approaches. Two of them require global clock synchronization. The first one minimizes the maximum TTR, while the second attempts of guarantying the even distribution of the rendezvous points in terms of time and channels. The third approach is an asynchronous variance of QCH, which guarantees rendezvous without the need of global clock synchronization. The protocols in [21] did not consider the different channel views and channel heterogeneity. The grid-based quorum rendezvous has been proposed in [21, 22]. We note that the grid quorum concept was not previously utilized in the context of channel allocation. In [21, 22], the authors proposed two different approaches to arrange users into grids. The first approach is Grid-Pair-on-Pair (POP). This grid cannot satisfy the rotation closure property (RCP). Another approach is the Grid-Diagonal (Diag), where the channel numbers are ordered according to a positive diagonal rule. The proposed algorithm in [22] used for nodes having asymmetric channel views, where the grid size can be adaptive according to the traffic load. Each node in the system maps its channel based on the channel quality without any information exchange. The authors in [2] assumed that each node keeps a list of channels, which is ordered according to their link quality. However, both works in [21, 22] used two different methods to map channels to time slots.

In group-based CCC schemes SUs usually observe the same spectrum availability in the vicinity. Common available channel in each group of SUs is allocated as control channel. This scheme facilitates the broadcast of control message within a group. Clustering scheme [23, 24] is a popular grouping method in distributed wireless networks. SUs are split into cluster according to cluster formulation algorithms. In the cluster technique there is a central entity to coordinate between SUs, which is cluster-head (CH). The CCC of a cluster is the common available channel to all cluster members, which is selected by the CH.

In [23], for each cluster, the CCC is selected by the CH in one-hop cluster framework. The formation of the cluster is based on the neighbor discovery by both channel scanning and then beacon broadcasting. The authors in [24] proposed a cluster-based structure, which allocates various channels for control at different clusters in the network. Accordingly, the control channel will be dynamically allocated based on PR activity. They use clustering to solve the problem of control channel assignment due to the spectrum heterogeneity and to solve the inter-cluster coordination and control channel migration problems.

## 1.3 Motivation:

One of the most challenging problems in CRNs is how to specify a CCC by SUs that enable them to establish connection and exchange control information. Rendezvous is an essential process between any two pair of nodes to enable communication. Note that without a successful rendezvous, the data communication, information exchange and spectrum management will be impossible [18].

In CRNs, the available channels can be determined by periodically sense spectrum of PUs by SUs. However, users have different geographic locations, which lead to different sensing results by SUs and so different sets of available channels for SUs (i.e., set of spectrum holes). A channel is available to SU if he can operate in this channel without interfering with the PUs. It is known that in CRN the sensing time and sensing capability are limited because of the dynamic of channel availability, which limits the spectrum sensing scope of SUs. If all SUs have the same set of available channels, then the model is said to be a symmetric. Otherwise, the network is called asymmetric model i.e. (different available channels for different users).

Several RDV approaches have been proposed in literature. Most of the rendezvous approaches were based on channel-hopping (CH) techniques because of its applicability in several scenarios. In CH technique, to achieve rendezvous between SUs over any of the available channels, each SU visits the channels in the network following a predefined strategy. In this technique, the time is divided into time slots and each SU hops among the set of available channels according to a predefined sequence. A rendezvous is guaranteed if two SUs hop to the same available channel at the same time slot [6], [15] and [25-28]. TTR is the most important requirement that should take into account to evaluate the performance of CH protocols, which indicates the number of hopping slots (time) that a SU takes to achieve a rendezvous. In addition, the maximum TTR (MTTR) and ETTR are two important criteria that are commonly used to evaluate the rendezvous performance. ETTR and MTTR respectively indicate the average and worst TTR scenario for guaranteeing rendezvous. If the MTTR of CH rendezvous algorithm is finite, then a rendezvous is guaranteed. To reduce channel access delay, the TTR should be small and bounded.

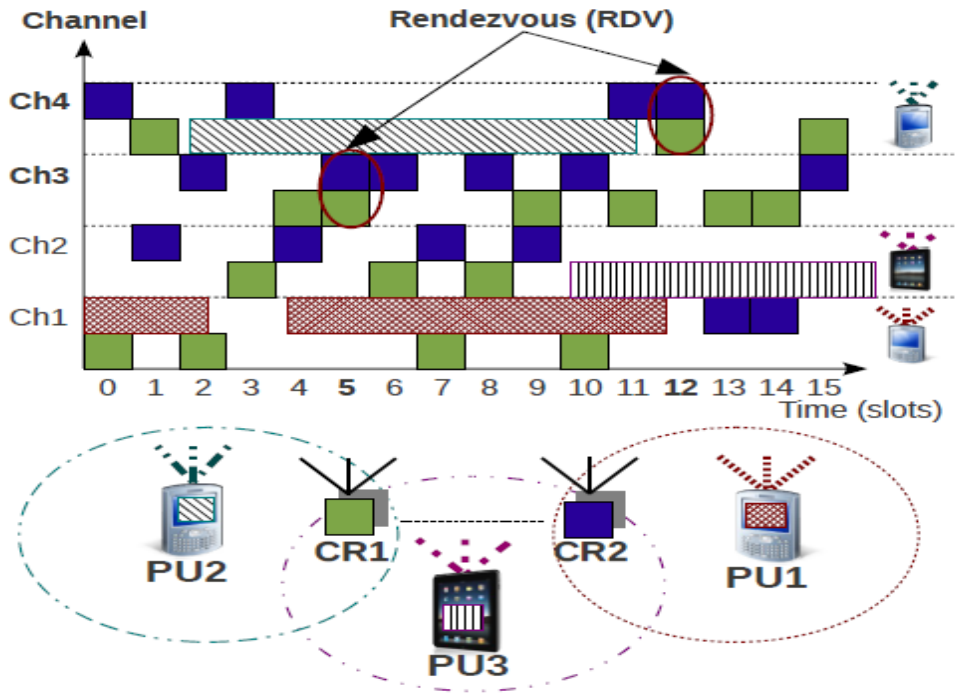

Fig.3: Rendezvous problem [29].

In Figure 3, PU1 is using channel 1 (neighbor of CR2) and PU2 is using channel 4 (neighbor of CR1) most of the time. Channel 2 is occupied by PU3 (where both CR1 and CR2 are within its transmission range). Within a 16 slots time period or cycle, CR1 and CR2 can meet twice (in time slot 5 on channel 3 and in time slot 12 on channel 4). We can see that CR users are never meet on channels 1 and 2. The situation in Figure 3 is very optimistic, where CR users are able to rendezvous twice. However, CR users might never meet when switching or hopping among channels is done randomly [29]. In this thesis, we deal with the aforementioned RDV problem by using one of the most promising rendezvous techniques, which is frequency-hopping quorum system (QS) technique.

## 1.4 Contributions:

The proposed algorithm employs the concept of channel-hopping. Our channel hopping is based on the idea of quorum system, where each SU selects their channel hopping sequence and then hops from one channel to another accordingly. There have been several algorithms based on quorum system with the objective of overcoming the RDV problem, where most of these algorithms are based on complex methods of quorum system such as cyclic quorum system. In this thesis, we propose different distributed rendezvous algorithms based on grid-based-quorum system that attempts at reducing the time of rendezvous. The proposed algorithms increase the probability of RDV within a single cycle by allowing nodes to meet on more than one channel according to intersection probability of quorum system. Our main proposed algorithm is called "Adaptive Quorum-based Channel-hopping Distribution Coordination Scheme for CRNs ", which is evaluated in homogeneous channel views to guarantee RDV on all available common channels. This means that all CR nodes are within the same quorum system and same transmission range.

In our proposed algorithm, the size of grid should be the same for all CR users which is N= $n \times n$, where N is the number of all PR channels and each user will select row and column from this grid. The main idea of our algorithm is to dynamically adjust the selection of the number of rows and columns for each node in the CRN according to the varying traffic loads such that the performance is improved in terms of number of RDV, TTR and energy consumption. This idea allows different CR users to dynamically select appropriate number of rows and columns from the quorum system based on the traffic load and network requirements in terms of the probability of RDV and energy consumption. The node can meet on more than one channel in a given cycle because of the unpredictable PU appearance. The proposed algorithms decrease TTR and increase the probability of RDV per cycle. The proposed adaptive algorithm differs from previous works (e.g. [5, 6]) in its adaption of cyclic different quorum systems. Our algorithm also differs from [18, 19] in its adaptive selection of the number of selected rows and columns in the grid-quorum system.

# Chapter 2: The Proposed Coordination Scheme:

In this thesis, we present new channel-hopping-based distributed rendezvous algorithms based on grid-based-quorum techniques. By using quorum systems, we can determine in which slots each channel should be visited. Based on intersection propriety of QS, we can guarantee RDV between nodes during a single cycle when the quorums are selected from the same quorum system. Each cycle period depends on the number of all PR channels N, in which each user will select number of rows and columns from $n \times n$ grid. Our algorithms reduce the TTR and increase the probability of RDV per cycle in an energy efficient way by dynamically adjusting the selection of the number of rows and columns for each node in the CRN according to the traffic load condition.

## 2.1 System Model:

### 2.1.1 CR Network Model

In this thesis, we consider a single-hop CRN that coexists with a number of PUs, where all the SUs are within the same coverage area and have the same set of channels. We assume that the CRN contains N licensed channels where $(i = 1,2, \ldots N)$, we also assume that the time T is divided into equal time slots t, where each time slot is enough to exchange the needed control messages. When two nodes want to rendezvous, they can follow 802.11 3-way handshake (RTS/CTS/DTS) process as shown in Figure 4. At the beginning of each timeslot, the CR senses the channel to detect the PU signals and avoid transmitting on this channel if PU signals are detected to avoid collision with PUs. After sensing the medium, the users select random backoff duration to avoid collision between SU. Then, the RTS/CTS/DTS can be exchanged which are needed for any CRN MAC design [40]. We note here that MAC design for CRNs is an important issue, but outside the scope of this thesis. We also assume that all nodes in the network are synchronized, where in CRN it is not easy to achieve time synchronization between nodes. Fortunately, various time synchronization protocols were proposed in literate for CRNs [38, 39]. Each SU in the CRN has two half-duplex transceiver, one for control exchange to switch between the N PR channels dynamically and the other for data transmissions.

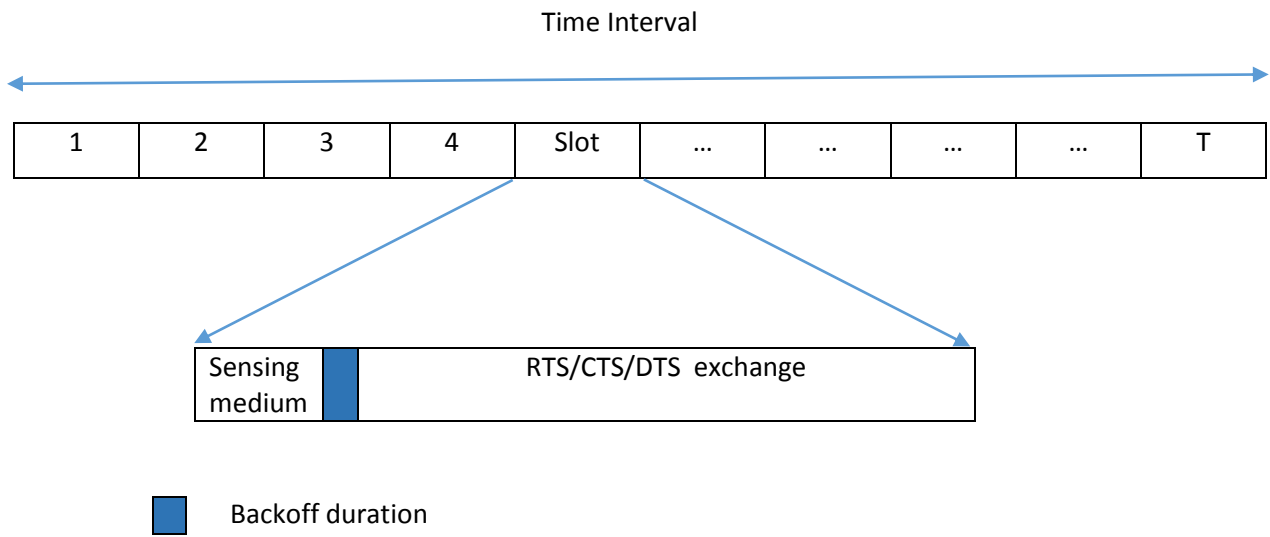

Fig.4: Time slot description.

## 2.1.2 The PR Activity Model

In each time slot, we assume the activity of PUs follows the two-state (ON-OFF) Markov model, where the ON period indicates the presence of the PU ( PR channel is busy) and the OFF period indicates that the channel is idle. The communication between the PUs are synchronized and the SUs which are also synchronized with the PUs can opportunistically access the available channels. Figure 5 shows the channel state for the ith channel. The PR activities indicate the distribution of the ON-OFF state, where in this model the ON-OFF states are independent random variables for a given channel. The average idle and busy periods for a given channel i are $\alpha_i$ and $\beta_i$, respectively where (1 ≤ i ≤ N), (e.g., Figures 5 and 6). Then, the idle and busy probabilities for the channel i are given by $P_I = \frac{\alpha_i}{\beta_i + \alpha_i}, 1 \leq i \leq N$ , $P_B = 1 - P_I$

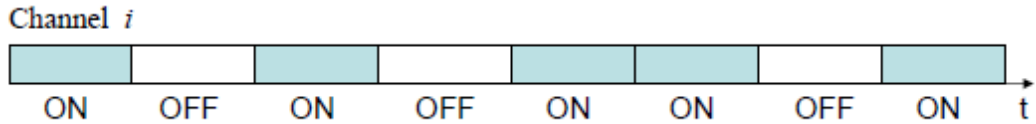

Fig.5: The channel occupancy for the ith channel by PUs [42].

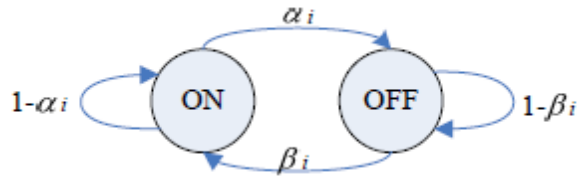

Fig.6: ON-OFF state channel availability model of a given PR channel [42].

### 2.1.3 Quorum System:

Quorum systems were primarily developed (and vastly used) in the field of operating systems [30]. Recently, wireless communications also have used quorum systems. The concept of a quorum has been used in achieving the distribution mutual exclusion, and widely used for a consistent data replication and agreement problem [9]. A quorum is a requested set that if it granted the permission, it will enable some actions. A quorum system has some properties (like intersection property) that can be utilized to establish communications without the need of a CCC, and hence overcome the RDV problem in CRNs. With quorum system, any two nodes can establish a connection as the overlap between any two CH sequences for two different users is guaranteed. Also, the use of quorum system can reduce the channel access delay, where the TTR between any pair of CH sequences has an upper bound (quorum system guarantees that within one period of time, there are more than one rendezvous between any pair of CH sequences). For more information on quorum, we refer the reader to the works in [8], [11].

We note here that for a network that contains of a set of M sites $\{P_1, P_2, ..., P_M\}$. In [8], the authors proposed an algorithm, where one of the consideration of this algorithm is to design a group of subsets $\{C_1, C_2, ..., C_M\}$, where $C_i$ is consist of $P_i$, for all $i \in 1,2,....M$. A Non empty intersection property is explained when $C_i \cap C_j \neq \emptyset$, for all $i \in 1,2,....M$. An equal work property is explained $|C_i| = k$, for all $i \in 1,2,....M$; where k< M. An equal responsibility property which say that $P_i$ is contained in $k$ $C_j$'s, for all $i \in 1,2,....M$. If all the above proprieties are satisfied, then a set of sites $C_i$ is called quorum.

## 2.2 The Grid-Based Quorum System

There are several types of quorum systems such as tours [34], tree-based [35], grid-based [9], and others [36, 37]. One of these types, which was proposed by Maekawa [9], is the grid-based system. In this system, a grid is used (in the shape of a square matrix) to logically organize the sites (elements) as $n \times n$ array. The union of full row and full column for a requesting site (elements in the array) is a quorum.

This system was vastly utilized in power-saving (PS) protocols for wireless networks. According, the beacon interval that the host should be awake it is referred to a quorum. In this case, the meeting of any two hosts at some beacon interval is guaranteed due to the quorum's properties. In grid-based quorum, the beacon intervals of PS nodes can be divided into groups, which include $n^2$ consecutive intervals, called a quorum group. In this system, each group of $n^2$ interval is organized in $n \times n$ grid matrix, where $n$ is a global parameter. From this grid, one row and one column are arbitrary picked for each node (i.e., this concept is shown in the Figure 7). For example, host A selects a quorum by picking row $R_a$ and column $C_a$, while host B selects row $R_b$ and column $C_b$. It can be noted that there are two intersections between hosts A and B. The union of these rows and columns represents the quorum intervals and the remaining intervals represent the non-quorum intervals. Each host in this system is awake for $\frac{2n-1}{n^2}$ intervals from the $n \times n$ grid, which lead to power saving by reducing the awake intervals.

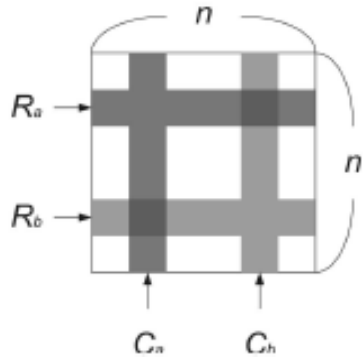

Fig.7: Grid-based quorum example [10].

Figure 8 shows an example of 16 consecutive beaconing intervals represented by a $4 \times 4$ grid for three nodes (A, B, C). Each of these three hosts selects randomly one row and one column randomly as its quorum interval, in which the host stays awake. For the entire beaconing interval, a host can sleep in non-quorum interval, in this example it is assumed that all hosts are synchronized. Host A selects its quorum intervals which are first row and first column, host B picks the second row and the third column, while host C selects the third row and the second column. This means that the beaconing intervals in which node A will be awake are 0,1,2,3,4,8 and 12, while the beacon intervals of node B are 2,4,5,6,7,10 and 14, and node C will be awake at the beaconing intervals 1,5,8,9,10,11 and 13. The intersection between hosts A and B occurs at beacon intervals 2 and 4, when both host are awake, while the intersection between hosts A and C occurs at beaconing intervals 1 and 8, while the intersection between hosts B and C occurs at beaconing intervals 5 and 10. It is clear that there are at least two overlaps between the intervals of any two nodes [21].

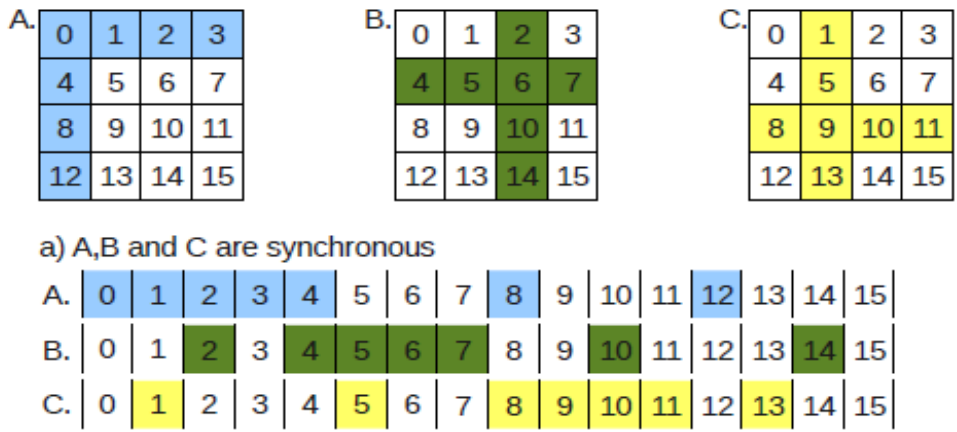

Fig.8: Grid-based quorum [21].

In [10], the authors defined rules to form a legal grid that called grid allocation rules. To keep the intersection property and guarantee that there are at least two intersections between any two SUs, we cannot arrange the grid randomly.

## 2.2.1 The Grid-Based Quorum Algorithm

In this thesis, we propose a distributed rendezvous algorithm that guarantees RDV during a single periods, where the period (cycle) contains a fixed number of a fixed-size time slots. Each cycle continuous $n^2$ time slots (called a group) and each group is arranged in $n \times n$ grid. Each time slot is mapped into one channel. So, the quorum system is used to guarantee RDV by determining which channel is mapped to each time slot. The quorum must be known for all users in the CRN. As mentioned before, we assume that all SUs are within the same coverage area and have the same number of channels. In this algorithm, the grid size depends on the number of all PR channels where $n = \lceil \sqrt{N} \rceil$. Then we map each channel to a grid index, where channel (1) is mapped to index 1, channel (2) is mapped to index 2, channel (3) is mapped to index (3), etc., (for example if there are 17 channels then $n = \lceil \sqrt{17} \rceil = 5$ and the grid size will be $5 \times 5$ then the channels will be assigned to slot in order as shown in Figure 9). All the nodes in the CRN have a complete knowledge of the index for each channel and for each user, where each channel will be mapped to one time slot.

| C1  | C2  | C3  | C4  | C5  |
|-----|-----|-----|-----|-----|
| C6  | C7  | C8  | C9  | C10 |
| C11 | C12 | C13 | C14 | C15 |
| C16 | C17 | C1  | C2  | C3  |
| C4  | C5  | C6  | C7  | C8  |

Fig.9: Grid-based Quorum with 17 channels.

In this algorithm, each sender and receiver select a number of rows and columns from the same grid quorum system to generate their channel hopping sequence separately. Thus, the rendezvous will be achieved within a sequence period on their common channels due to the intersection property of quorum system.

Consider two users that want to communicate, say users A and B. Recall that all the users in the CRN share the same quorum system. Assume that there are 16 channels in the network, where they are arranged in the grid quorum system. Each user will randomly select one row and one column from the grid-quorum system. The union of the row and column determines the quorum interval, where the users can exchange control information as shown in Figure 9. It is clear that users will intersect on at least two quorum slots. For example, users A and B will exchange control information in channels 7 and 10 and at the $7^{th}$ and $10^{th}$ quorum slots.

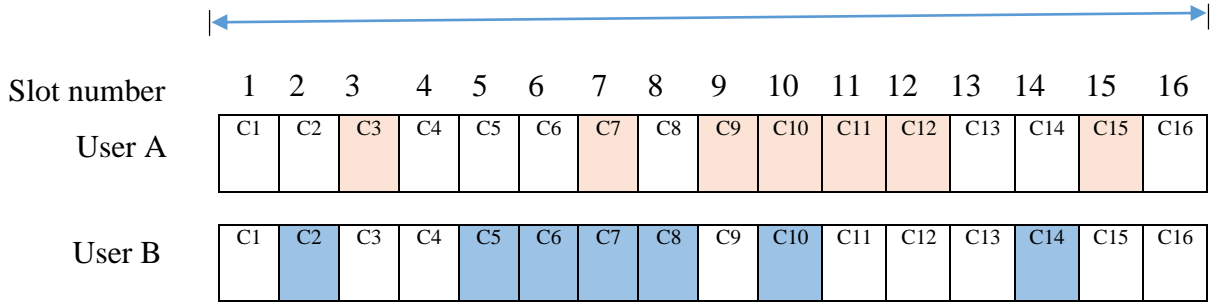

Fig.10: Grid-based Quorum with 16 channels.

## 2.2.2 The Proposed Design Variants

We propose three different algorithms to choose the rows and columns in the quorum system.

### 2.2.2.1 The Proposed 1×1-Qourum RDV Scheme:

The first algorithm is based on random selection, where each SU randomly selects one row and one column (1×1-selection) from the quorum system and this results in 2n-1 quorum intervals. It is worth mentions that this variant was previously proposed in [21, 22] with different channel mapping procedure. Given two CR nodes, which are perfectly synchronized, we can show that there will be at least two rendezvous in two channels in each cycle.

Case 1: Selecting different rows and different columns:

| 1 | 2 | ................................. | n-1 | n |
|---|---|---|---|---|
| n+1 | n+2 | ................................. | 2n-1 | 2n |
| . | . | . | . | . |
| . | . | . | . | . |
| . | . | . | . | . |
| . | . | . | . | . |
| . | . | . | . | . |
| . | . | . | . | . |
| . | . | . | . | . |
| . | . | . | . | . |
| $n^2 - n + 1$ | $n^2 - n + 2$ | ................................. | $n^2 - 1$ | $n^2$ |

Fig.11: Grid-based quorum (1×1-selection) case 1.

By inspection, it can be noted that there are at least two possible RDVs.

Now, given the states per P$_I$ (idle probability) the average number of RDV per time slot is given by $\frac{2}{n^2} \times$P$_I$

Case 2: Selecting different rows and same column or same row and different columns: In this case, the maximum number of possible RDVs is n, while the average number of RDVs per time slot in case of PR activities is given by $\frac{n}{n^2} \times P_I$.

Fig.12: Grid-based quorum (1×1-selection) case 2.

Case 3: Selecting the same column and the same row: In this case, 2n-1 number of possible RDVs, while the average number of RDVs per time slot in case of PR activities is given by $\frac{2n-1}{n^2} \times P_I$.

Fig.13: Grid-based quorum (1×1-selection) case 3.

### 2.2.2.2 The Proposed Fixed N×M-Quorum RDV Scheme:

The second algorithm is based on selecting variable number of rows and columns. In this method, each user determines the appropriate number of rows and columns that a user can select such that the quorum conditions are met. In this method, we study two states. The first one is when users select 2 rows and 1 column (2×1-selection) and the second one, represents the case when users select 2 rows and 2 columns (2×2-selection).

A. Selecting 2 rows and 1 column

Case 1: Selecting different rows and different columns: In this case, the maximum number of possible RDVs is 4, while the average number of RDVs per time slot in case of PR activities is given by $\frac{4}{n^2} \times P_I$.

| 1 | 2 | ………………………………… | n-1 | n |
|---|---|---|---|---|
| n+1 | n+2 | ………………………………… | 2n-1 | 2n |
| 2n+1 | 2n+2 | ………………………………… | 3n-1 | 3n |
| . | . | . | . | . |
| . | . | . | . | . |
| . | . | . | . | . |
| . | . | . | . | . |
| . | . | . | . | . |
| . | . | . | . | . |
| . | . | . | . | . |
| $n^2-n+1$ | $n^2-n+2$ | ………………………………… | $n^2-1$ | $n^2$ |

Fig.14: Grid-based quorum (2×1-selection) case 1.

Case 2: Selecting one common row and different columns: In this case, the maximum number of possible RDVs is n+2, while the average number of RDVs per time slot is given by $\frac{n+2}{n^2} \times P_I$.

| 1 | 2 | ............ | n-1 | n |
|---|---|---|---|---|
| n+1 | n+2 | ............ | 2n-1 | 2n |
| 2n+1 | 2n+2 | ............ | 3n-1 | 3n |
| . | . | . | . | . |
| . | . | . | . | . |
| . | . | . | . | . |
| . | . | . | . | . |
| . | . | . | . | . |
| . | . | . | . | . |
| . | . | . | . | . |
| $n^2 - n + 1$ | $n^2 - n + 2$ | ............ | $n^2 - 1$ | $n^2$ |

Fig.15: Grid-based quorum (2×1-selection) case 2.

Case 3: Selecting the same row and different columns: In this case, the maximum number of possible RDVs is 2n, while the average number of RDVs per time slot is given by $\frac{2n}{n^2} \times P_I$.

| 1 | 2 | ............ | n-1 | n |
|---|---|---|---|---|
| n+1 | n+2 | ............ | 2n-1 | 2n |
| 2n+1 | 2n+2 | ............ | 3n-1 | 3n |
| . | . | . | . | . |
| . | . | . | . | . |
| . | . | . | . | . |
| . | . | . | . | . |
| . | . | . | . | . |
| . | . | . | . | . |
| . | . | . | . | . |
| $n^2 - n + 1$ | $n^2 - n + 2$ | ............ | $n^2 - 1$ | $n^2$ |

Fig.16: Grid-based quorum (2×1-selection) case 3.

Case 4: Selecting the same rows and columns: In this case, the maximum number of possible RDVs is 3n-2, while the average number of RDVs per time slot case is given by $\frac{3n-2}{n^2} \times P_I$.

| 1 | 2 | ............................................ | n-1 | n |
|---|---|---|---|---|
| n+1 | n+2 | ............................................ | 2n-1 | 2n |
| 2n+1 | 2n+2 | ............................................ | 3n-1 | 3n |
| . | . | . | . | . |
| . | . | . | . | . |
| . | . | . | . | . |
| . | . | . | . | . |
| . | . | . | . | . |
| . | . | . | . | . |
| . | . | . | . | . |
| $n^2 - n + 1$ | $n^2 - n + 2$ | ............................................ | $n^2 - 1$ | $n^2$ |

Fig.17: Grid-based quorum (2×1-selection) case 4.

B. Selecting 2 rows and 2 columns

Case 1: Selecting different rows and different columns: In this case, the maximum number of possible RDVs is 8, while the average number of RDVs per time slot is given by $\frac{8}{n^2} \times P_I$.

| 1 | 2 | ............................................ | n-1 | n |
|---|---|---|---|---|
| n+1 | n+2 | ............................................ | 2n-1 | 2n |
| 2n+1 | 2n+2 | ............................................ | 3n-1 | 3n |
| . | . | . | . | . |
| . | . | . | . | . |
| . | . | . | . | . |
| . | . | . | . | . |
| . | . | . | . | . |
| . | . | . | . | . |
| . | . | . | . | . |
| $n^2 - n + 1$ | $n^2 - n + 2$ | ............................................ | $n^2 - 1$ | $n^2$ |

Fig.18: Grid-based quorum (2×2-selection) case 1.

Case 2: one common row and different column selection: In this case, the maximum number of possible RDVs is n+4, while the average number of RDVs per time slot is given by $\frac{n+4}{n^2} \times P_I$.

| 1 | 2 | ............................................. | n-1 | n |
|---|---|---|---|---|
| n+1 | n+2 | ............................................. | 2n-1 | 2n |
| 2n+1 | 2n+2 | ............................................. | 3n-1 | 3n |
| . | . | . | . | . |
| . | . | . | . | . |
| . | . | . | . | . |
| . | . | . | . | . |
| . | . | . | . | . |
| . | . | . | . | . |
| . | . | . | . | . |
| $n^2 - n + 1$ | $n^2 - n + 2$ | ............................................. | $n^2 - 1$ | $n^2$ |

Fig.19: Grid-based quorum (2×2-selection) case 2.

Case 3: Selecting common rows and different columns: In this case, the maximum number of possible RDVs is 2n, while the average number of RDVs per time slot is given by $\frac{2n}{n^2} \times P_I$.

| 1 | 2 | ............................................. | n-1 | n |
|---|---|---|---|---|
| n+1 | n+2 | ............................................. | 2n-1 | 2n |
| 2n+1 | 2n+2 | ............................................. | 3n-1 | 3n |
| . | . | . | . | . |
| . | . | . | . | . |
| . | . | . | . | . |
| . | . | . | . | . |
| . | . | . | . | . |
| . | . | . | . | . |
| . | . | . | . | . |
| $n^2 - n + 1$ | $n^2 - n + 2$ | ............................................. | $n^2 - 1$ | $n^2$ |

Fig.20: Grid-based quorum (2×2-selection) case 3.

Case 4: Selecting one common row and one common column: In this case, the maximum number of possible RDVs is 2n+1, while the average number of RDVs per time slot is given by $\frac{2n+1}{n^2} \times P_I$.

| 1 | 2 | ................................ | n-1 | n |
|---|---|---|---|---|
| n+1 | n+2 | ................................ | 2n-1 | 2n |
| 2n+1 | 2n+2 | ................................ | 3n-1 | 3n |
| . | . | . | . | . |
| . | . | . | . | . |
| . | . | . | . | . |
| $n^2-n+1$ | $n^2-n+2$ | ................................ | $n^2-1$ | $n^2$ |

Fig.21: Grid-based quorum (2×2-selection) case 4.

Case 5: Selecting one common row and common column: In this case, the maximum number of possible RDVs is 3n-2, while the average number of RDVs per time slot is given by $\frac{3n-2}{n^2} \times P_I$

| 1 | 2 | ................................ | n-1 | n |
|---|---|---|---|---|
| n+1 | n+2 | ................................ | 2n-1 | 2n |
| 2n+1 | 2n+2 | ................................ | 3n-1 | 3n |
| . | . | . | . | . |
| . | . | . | . | . |
| . | . | . | . | . |
| $n^2-n+1$ | $n^2-n+2$ | ................................ | $n^2-1$ | $n^2$ |

Fig.22: Grid-based quorum (2×2-selection) case 5.

Case 6: Selecting common rows and common columns: In this case, the maximum number of possible RDVs is 4n-4, while the average number of RDVs per time slot is given by $\frac{4n-4}{n^2} \times P_I$.

| 1 | 2 | ............................................ | n-1 | n |
|---|---|---|---|---|
| n+1 | n+2 | ............................................ | 2n-1 | 2n |
| 2n+1 | 2n+2 | ............................................ | 3n-1 | 3n |
| . | . | . | . | . |
| . | . | . | . | . |
| . | . | . | . | . |
| . | . | . | . | . |
| . | . | . | . | . |
| . | . | . | . | . |
| . | . | . | . | . |
| . | . | . | . | . |
| $n^2 - n + 1$ | $n^2 - n + 2$ | ............................................ | $n^2 - 1$ | $n^2$ |

Fig.22: Grid-based quorum (2×2-selection) case 6.

### 2.2.2.3 The Proposed Adaptive N×M-Quorum RDV Scheme:

We also propose an adaptive algorithm that selects dynamic number of rows and columns, where the selection of rows and columns will be adaptively based on the traffic load. We also enhance the previous algorithms by utilizing intersection proprieties in quorum system. We divide the traffic into three regions (low, moderate and high), where for each region we select different number of rows and columns. For low region, we adopt the 2×2-selection which consumes less energy with high average number of successful RDV. The best selection for the moderate region is the 2×1-selection, while for the high traffic region, we use the 1×1-selection, which provides comparable values of average number of successful RDV as the 2×2-selection with less amount of energy consumption.

> **Practical implementation**
>
> We assume traffic estimation mechanism [47] is in place to determine the various traffic regions (low, moderate and high) the details of such mechanism is outside the scope of our work and left for future work (please for more details referred to section 4.2).

### 2.2.3 Illustrative Example

We consider different examples to show the results of our different selection variants for N=16 channels and a grid size of 4×4.

Firs, we study 1 row and 1 column selection for all cases.

Case 1: Different rows and different columns selection: The RDV occurs on two channels 7 and 12, which are the minimum number of RDVs.

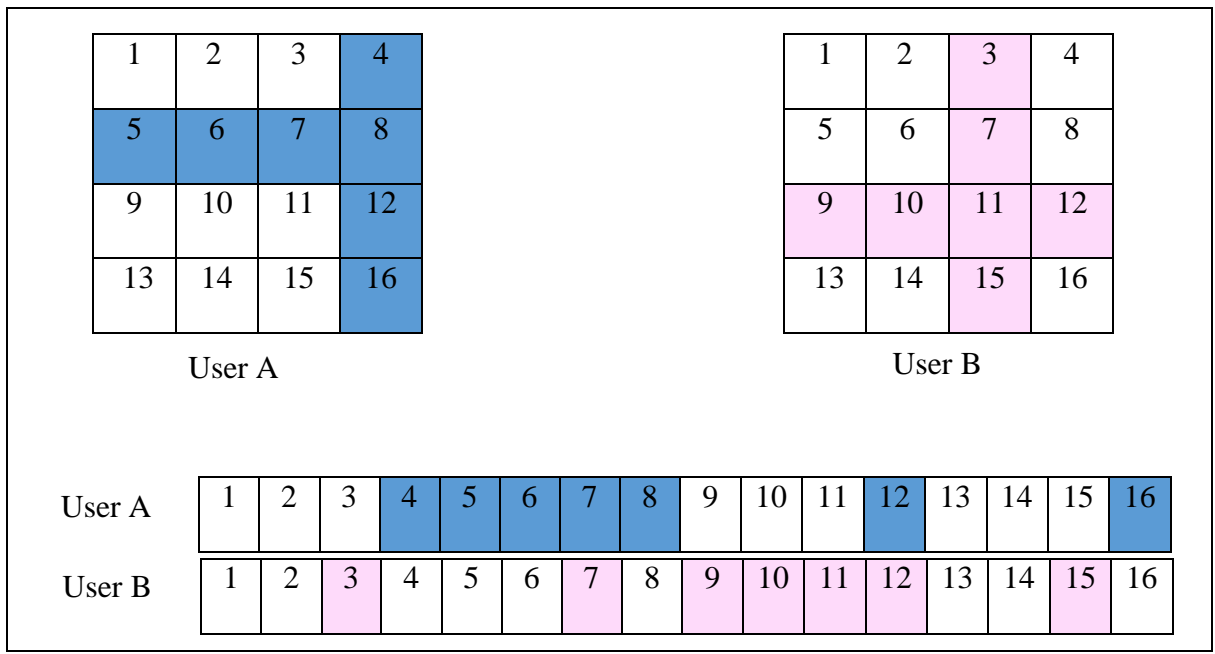

Fig.23: An example of 4×4 grid with N=16, 1 row and 1 column selection case 1.

Case 2: Different rows and same column or same row and different columns selection: We note that the RDVs occurs on channels 9, 10, 11 and 12. As mention before, the maximum number of RDVs is n = 4 in our example.

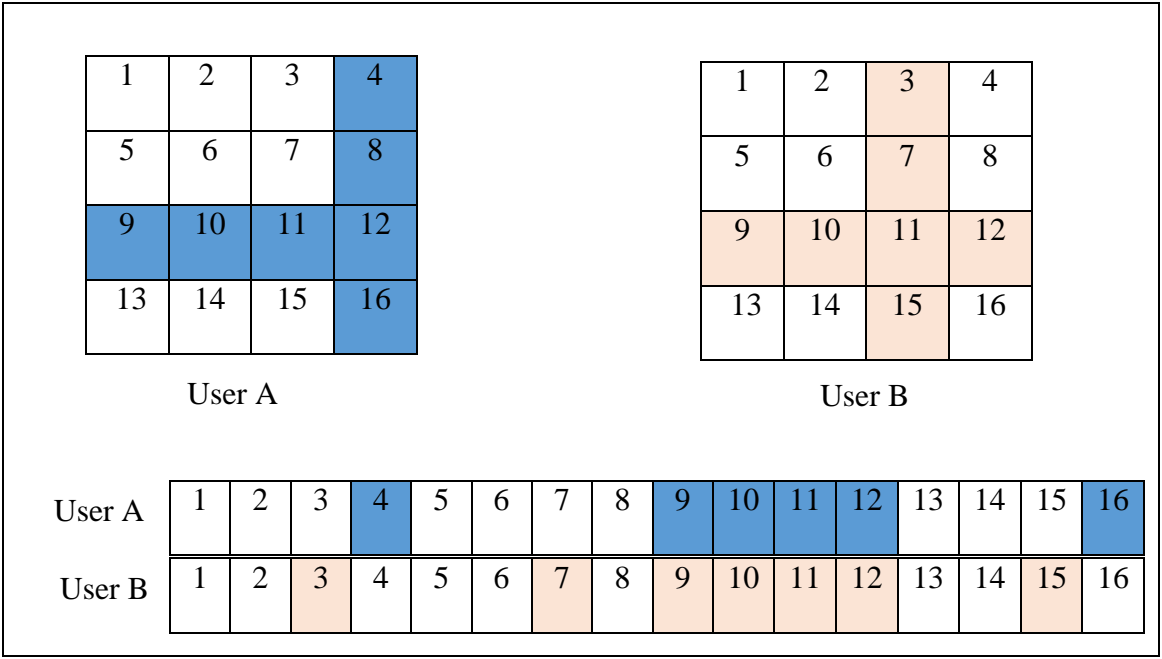

Fig.24: An example of 4×4 grid with N=16, 1 row and 1 column selection case 2.

Case 3: Selecting the same column and the same row: In this case, the maximum number of RDVs, is achieved, which is 2n-1=7. This occurs on channels 3, 7, 9,10,11,12 and 15.

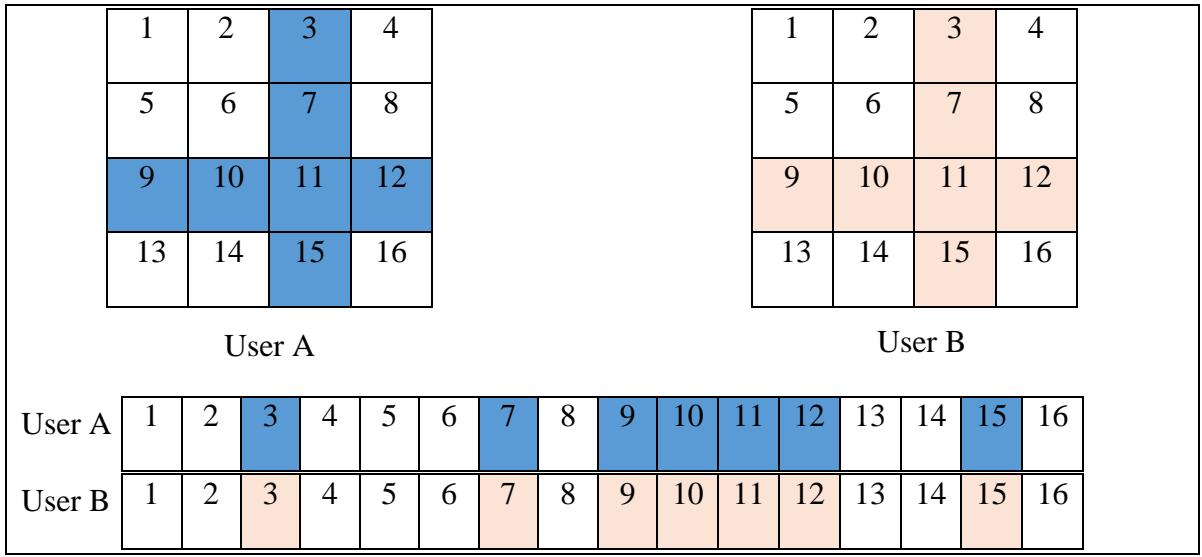

Fig.25: An example of 4×4 grid with N=16, 1 row and 1 column selection case 3.

Second, we study the variant when 2 rows and 1 column are selected.

Case 1: Different row and different column selection: The RDV occur on channels 4,7,12 and 1, which is the minimum number of RDVs in this variant.

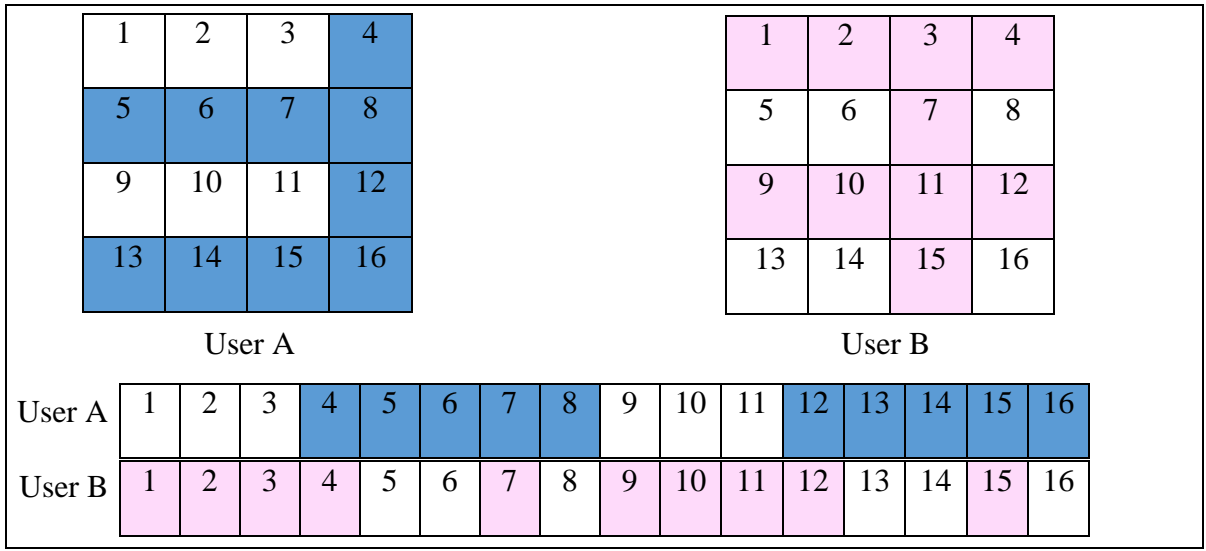

Fig.26: An example of 4×4 grid with N=16, 2 row and 1 column selection case 1.

Case 2: One common row and different column selection: The number of RDVs increases to 6 and occur on channels 1, 2, 3,4,12 and 15.

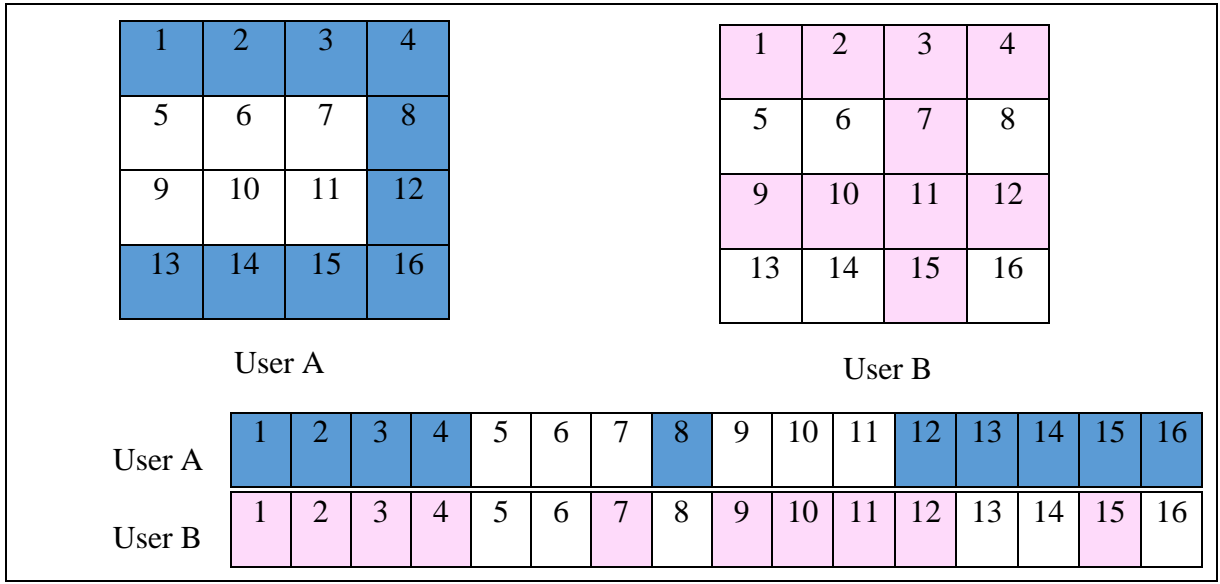

Fig.27: An example of 4×4 grid with N=16, 2 row and 1 column selection case 2.

Case 3: Same row and different column selection: the RDVs in this case occur on channels 1,2,3,4,13,14,15 and 16.

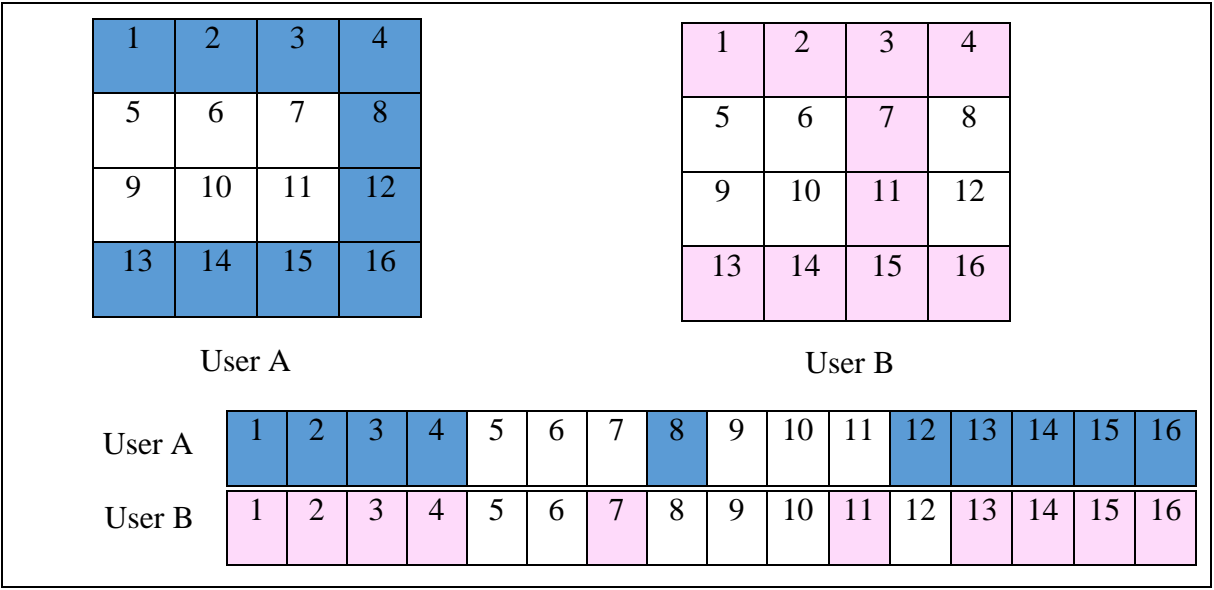

Fig.28: An example of 4×4 grid with N=16, 2 row and 1 column selection case 3.

Case 4: Same row and column selection: In this case, the maximum number of RDVs is achieved for this variant, which is 3n-2 and occur on channels 1,2,3,4,8,12,13,14,15 and 16.

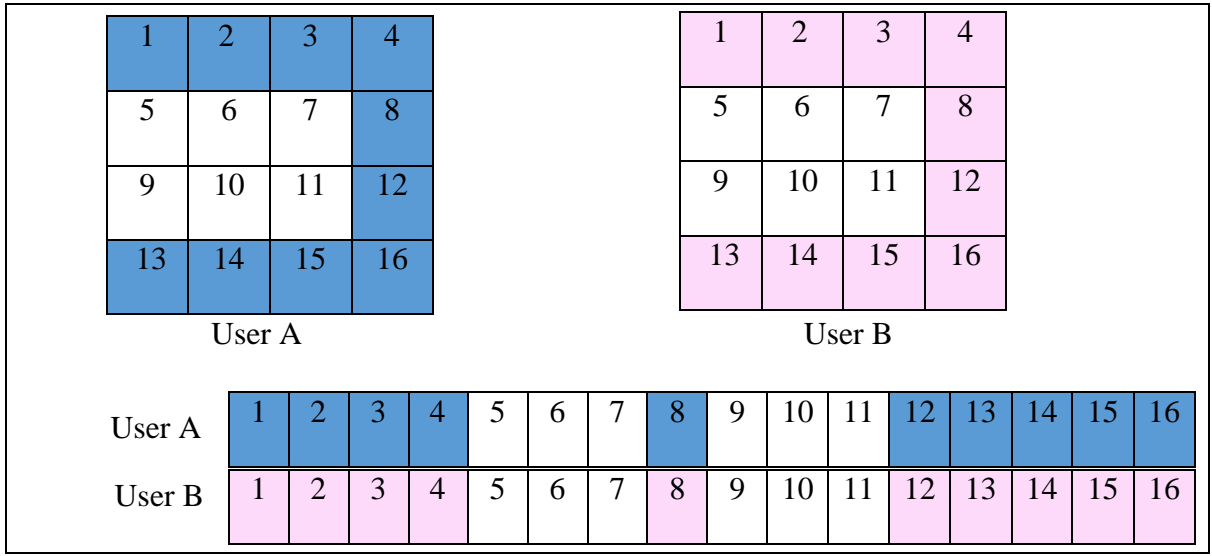

Fig.29: An example of 4×4 grid with N=16, 2 row and 1 column selection case 4.

The last example shows the variant when 2 rows and 2 columns are selected.

Case 1: Different rows and different columns selection: the RDVs occur on channels 2,4,5,7,10,12,13 and 15, which represented the minimum number in this variant and (the number of RDVs increases when selecting more rows and columns).

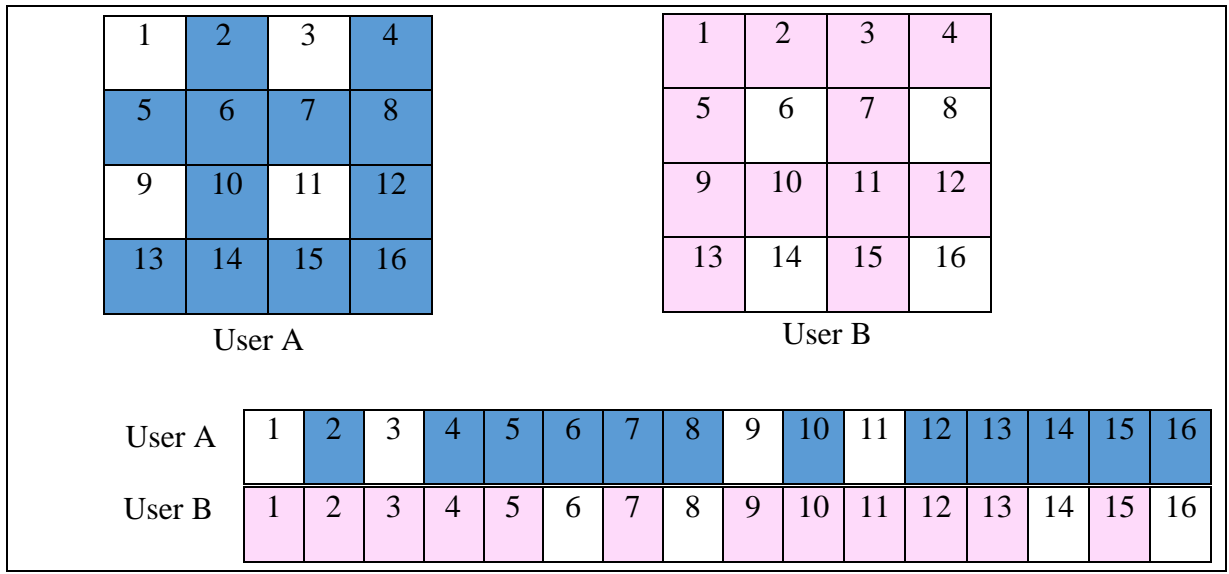

Fig.30: An example of 4×4 grid with N=16, 2 row and 2 column selection case 1.

Case 2: one common row and different column selection: the RDVs occur on channels 1,2,3,4,10,12,13 and 15.

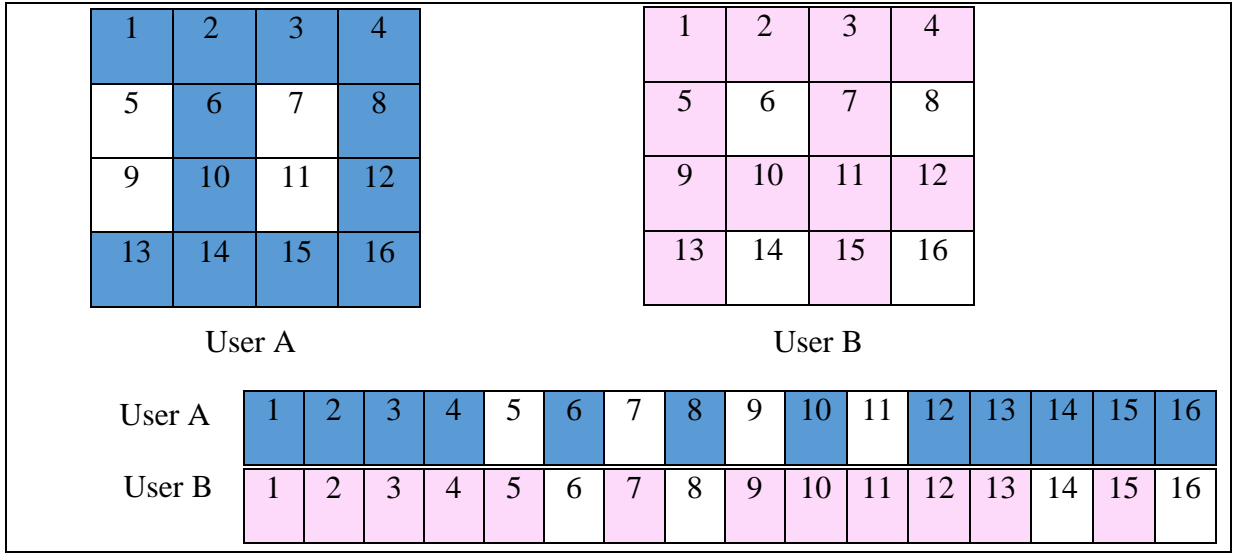

Fig.31: An example of 4×4 grid with N=16, 2 row and 2 column selection case 2.

Case 3: Common row and different columns selection: the RDVs occur on channels 1, 2, 3,4,9,10,11 and 13.

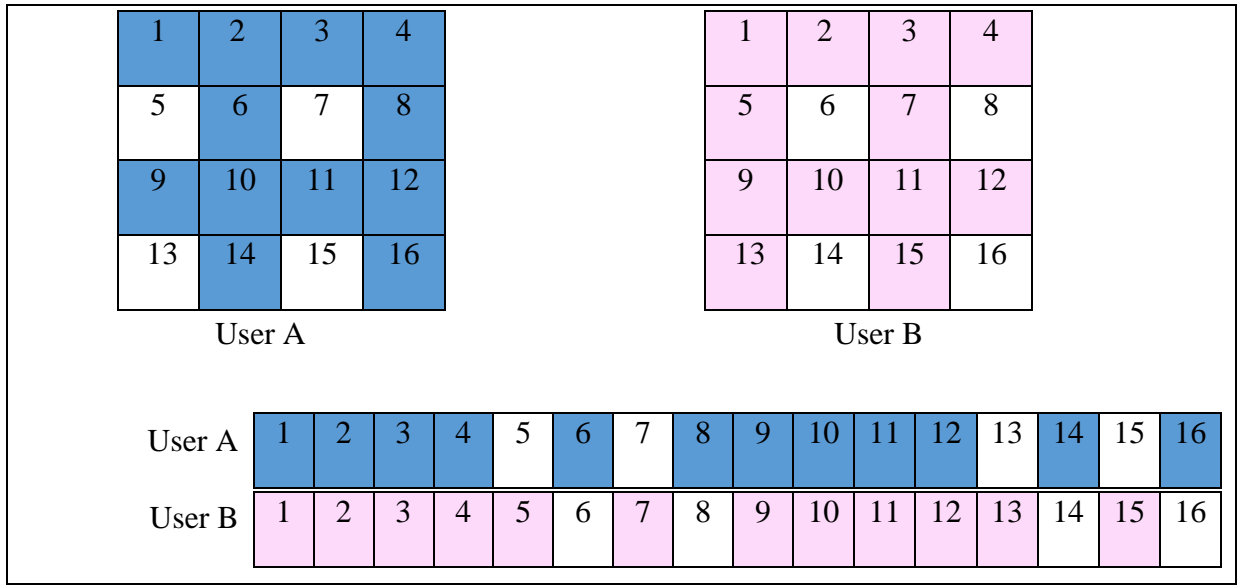

Fig.32: An example of 4×4 grid with N=16, 2 row and 2 column selection case 3.

Case 4: One common row and one common column selection: the RDVs occur on channels 1,2,3,4,5,9,12,13 and 15.

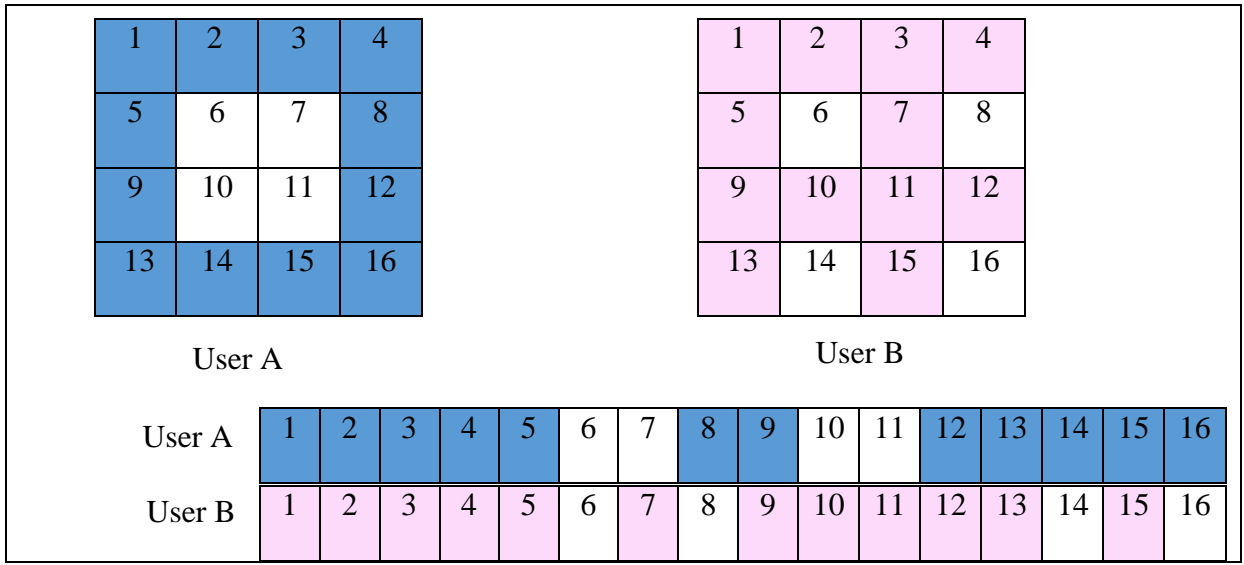

Fig.33: An example of 4×4 grid with N=16, 2 row and 2 column selection case 4.

Case 5: One common row and common columns selection: the RDVs occur on channels 1, 2,3,4,5,7,9,11,13 and 15.

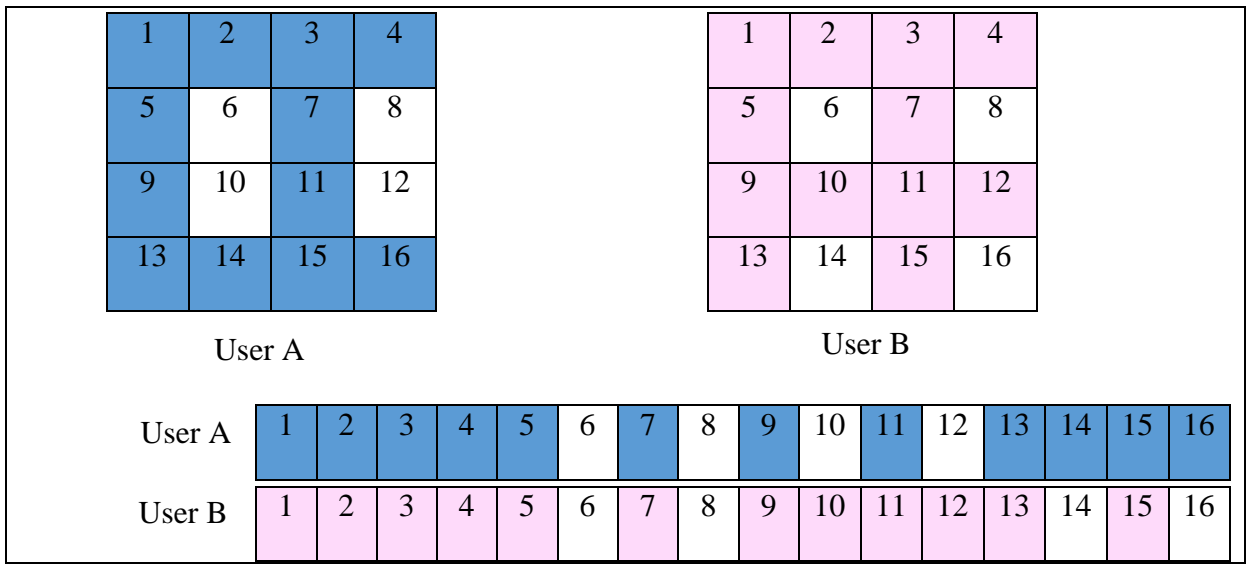

Fig.34: An example of 4×4 grid with N=16, 2 row and 2 column selection case 5.

Case 6: Common rows and common columns selection: in this case, the maximum number of RDVs can be achieved which occur on channels 1,2,3,4,5,7,9,10,11,12,13 and 15. We can observe that 2×2-selection provides the largest number of RDVs.

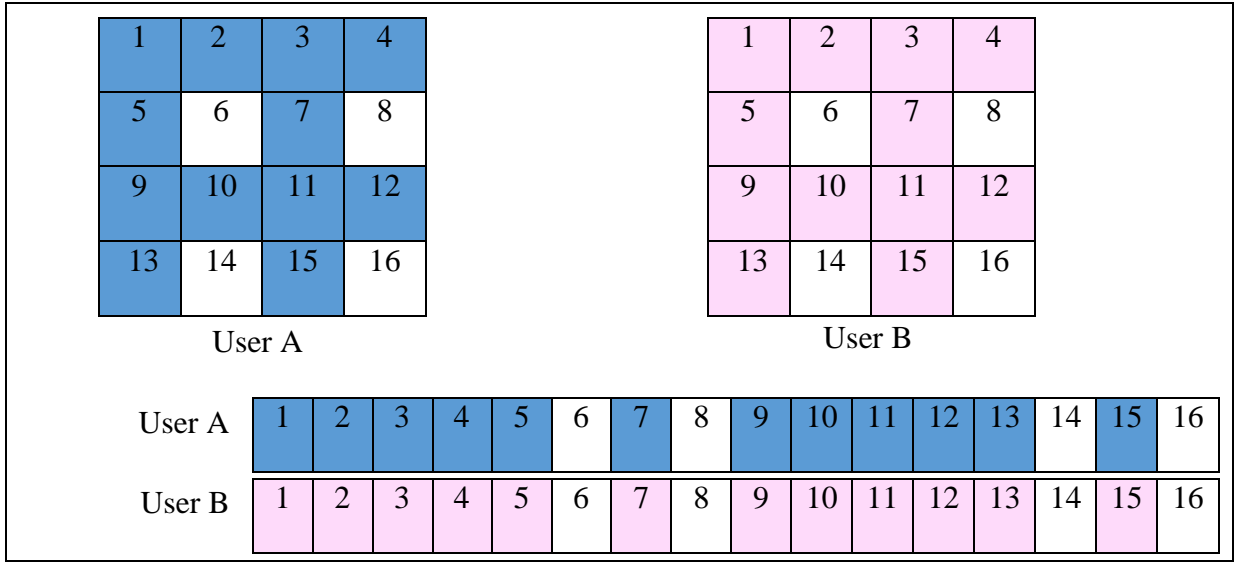

Fig.35: An example of 4×4 grid with N=16, 2 row and 2 column selection case 5.

Figure 36 shows the flow-chart of our proposed algorithm for the different algorithms.

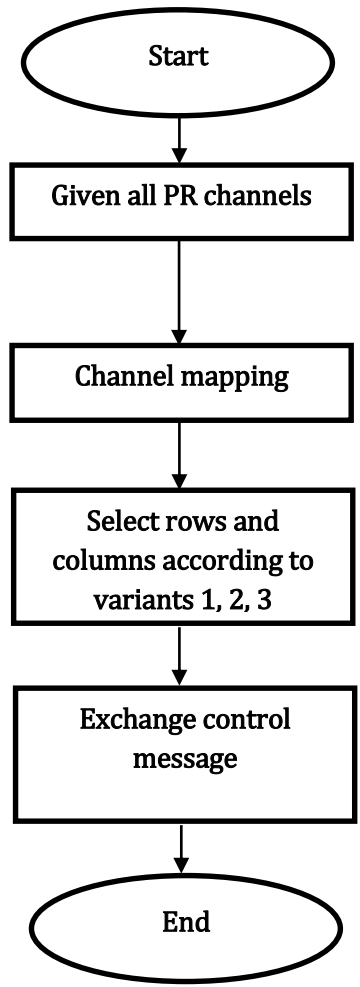

Fig.36: The Flowchart of our proposed algorithm.

Figure 37 shows the flow-chart of our proposed algorithm for the adaptive algorithms.

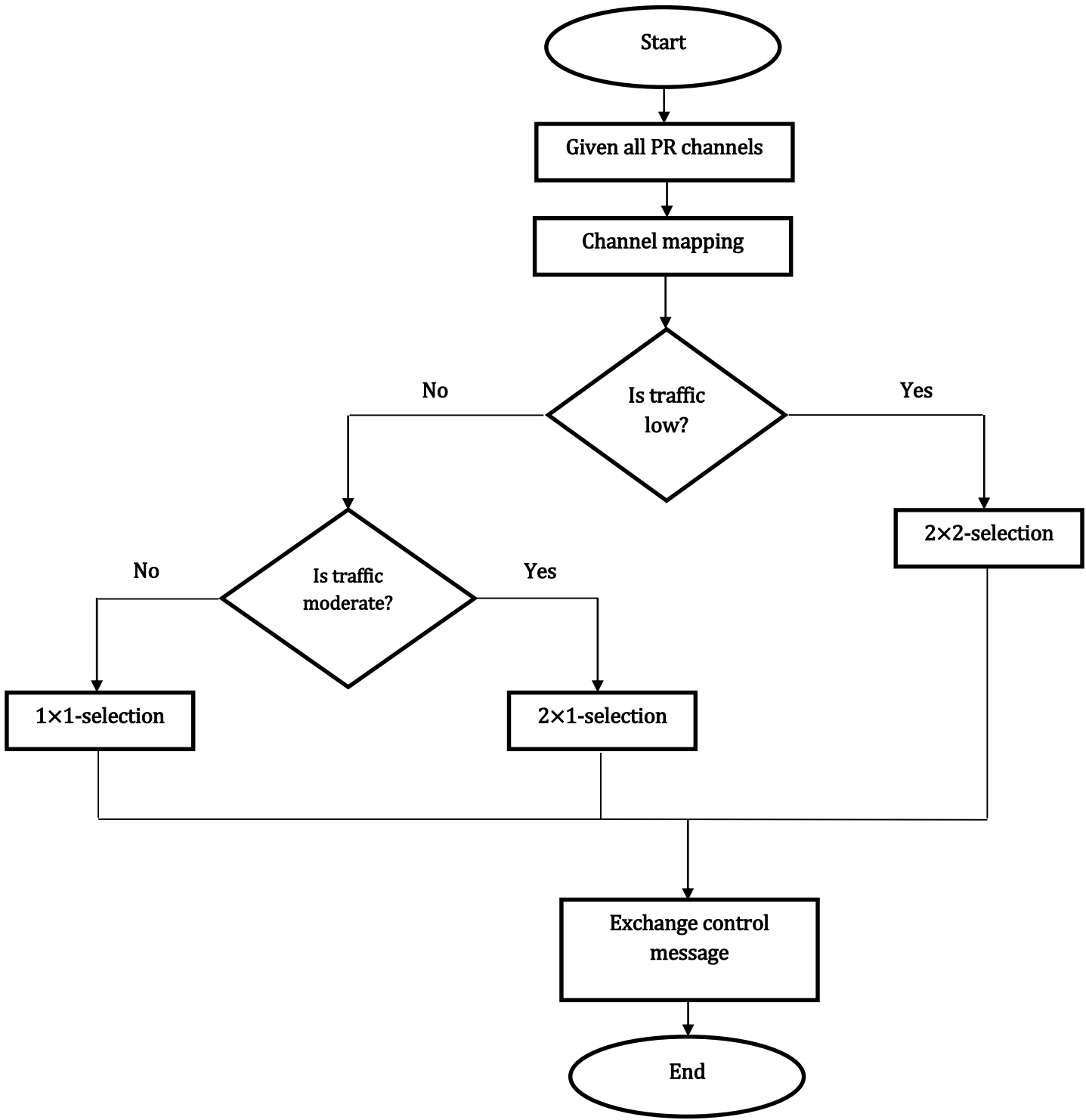

Fig.37: The Flowchart of our proposed adaptive algorithm.

# Chapter 3: Performance Evaluation:

## 3.1 Simulation Setup:

We analyze the proposed schemes using Matlab simulations. In our simulations, we consider that there are several PRNs which coexist with a CRN and there are no connection between them. In our simulations, we consider three distinct behavior of PRs activity when the probability of idle for each channels $P_I$=.1 (high activity), $P_I$=.5 (moderate activity), and low activity when $P_I$=.9. We test the performance of our schemes for (N=9, 16, 25) channels and for grid size $n \times n$. We consider that there are 26 or 50 transmitting CR pairs. At each time slot, the CR transmitter randomly selects its receiver. Each node can perform traffic estimation for the PUs and SUs. We model the availability of channels by using the 2-states Markov chain. The duration of each time slot is 0.384 ms, which represents the time needed for successful control packet exchanges (160 bits for RTS, 112 bits for CTS and 112 bits for DTS) with transmission control rate of 1 Mbps. Our results are averaged over 800000 time slots.

## 3.2 Simulation Results:

In our simulations, we use four metrics to compare the performance of our different schemes, i.e.

- The average number of successful rendezvous, which is defined as the average number of successful meeting per time slot [25].
- Average TTR, which is the number of time slots that two users need to wait on average before they found common channel to rendezvous [44].

- Normalized energy per successful RDV, which indicates the average number of active slots per a successful transmission.
- Forced blocking due to PR activity, which indicates the effect of PR activity on network performance [45].

For each metrics, we perform two experiments: we vary the number of SUs from 4 to 50 and study the effect of increasing the number of SUs for ($P_I$=0.1, 0.5 and 0.9). In the second experiment, we vary the idle probabilities from 0.1 to 1 and study the effect of increasing the idle probability for 26 and 50 SUs. Both of these experiments are evaluated for grid sizes $3 \times 3$, $4 \times 4$ and $5 \times 5$. The performance of the proposed variants (2×1-selection, 2×2-selection and adaptive selection) is compared with a reference scheme that uses only 1×1-selection [21, 22].

### 3.2.1 Average Number of Successful Rendezvous:

We first study the average number of rendezvous per time slot, which is defined as the percentage of successful meeting for all nodes per time slots. We evaluate the average number of rendezvous as a function of $P_I$ for different number of CR users. As show in Figures 38-40, in respective of the grid size, the average number of rendezvous for all schemes increases as $P_I$ increases due to the fact that increasing the number of available channels increases the number of successful meeting on more channels.

Figure 38 shows the average number of rendezvous for grid size=3×3 and 9 channels. In Figure 38, 2×2-selection outperforms the other two schemes (up to 100% improvement in the average number of rendezvous).

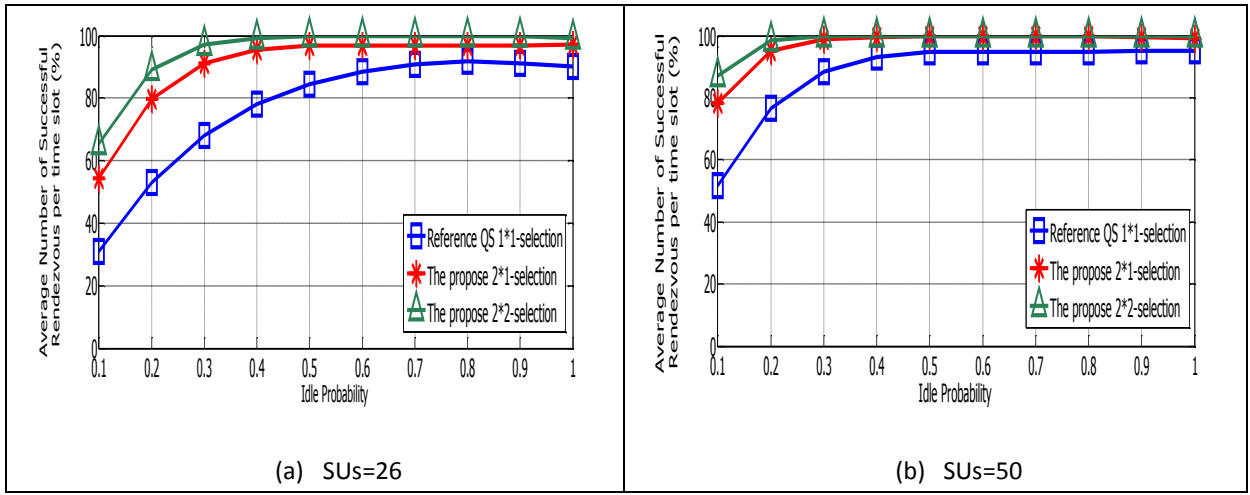

Fig.38: Average number of successful rendezvous per time slot vs. $P_I$, for grid size=3×3.

Next, we consider the average number of rendezvous for grid size of 4×4 and 16 channels. Figure 39 depicts similar behavior to that of grid size=3×3, but the maximum average rendezvous is around 85% to 100% for SUs=26 and 90% to 100% for SUs=50.

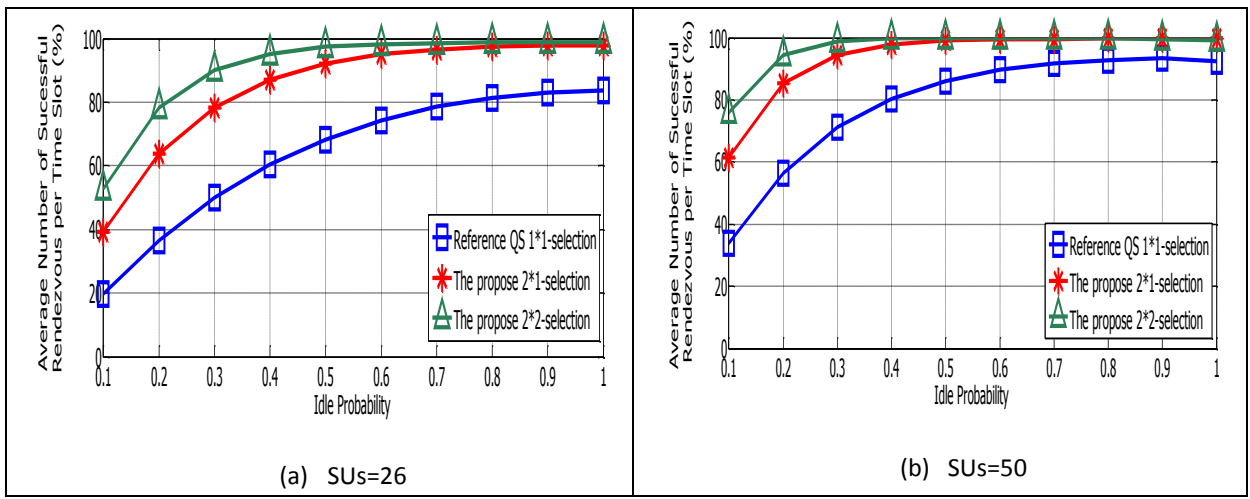

Fig 39: Average number of successful rendezvous per time slot vs. $P_I$, for grid size=4×4.

Figure 40 shows similar behavior to that in Figure 38 and 39, but with maximum average rendezvous of 73% to 95% for SUs=26 and 84% to 97% for SUs=50.

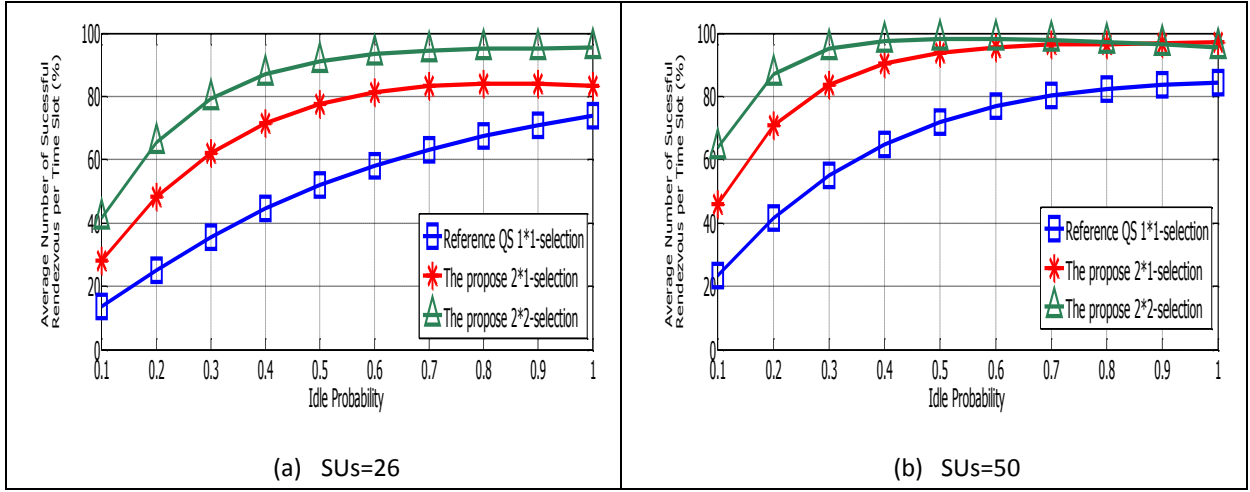

(a) SUs=26  (b) SUs=50

Fig.40: Average number of successful rendezvous per time slot vs. $P_I$, for grid size=5×5.

Second, we study the average number of RDVs as a function of the number of SUs for different values of $P_I$ while varying the grid sizes and the number of channels in Figure 41 and 42, respectively.

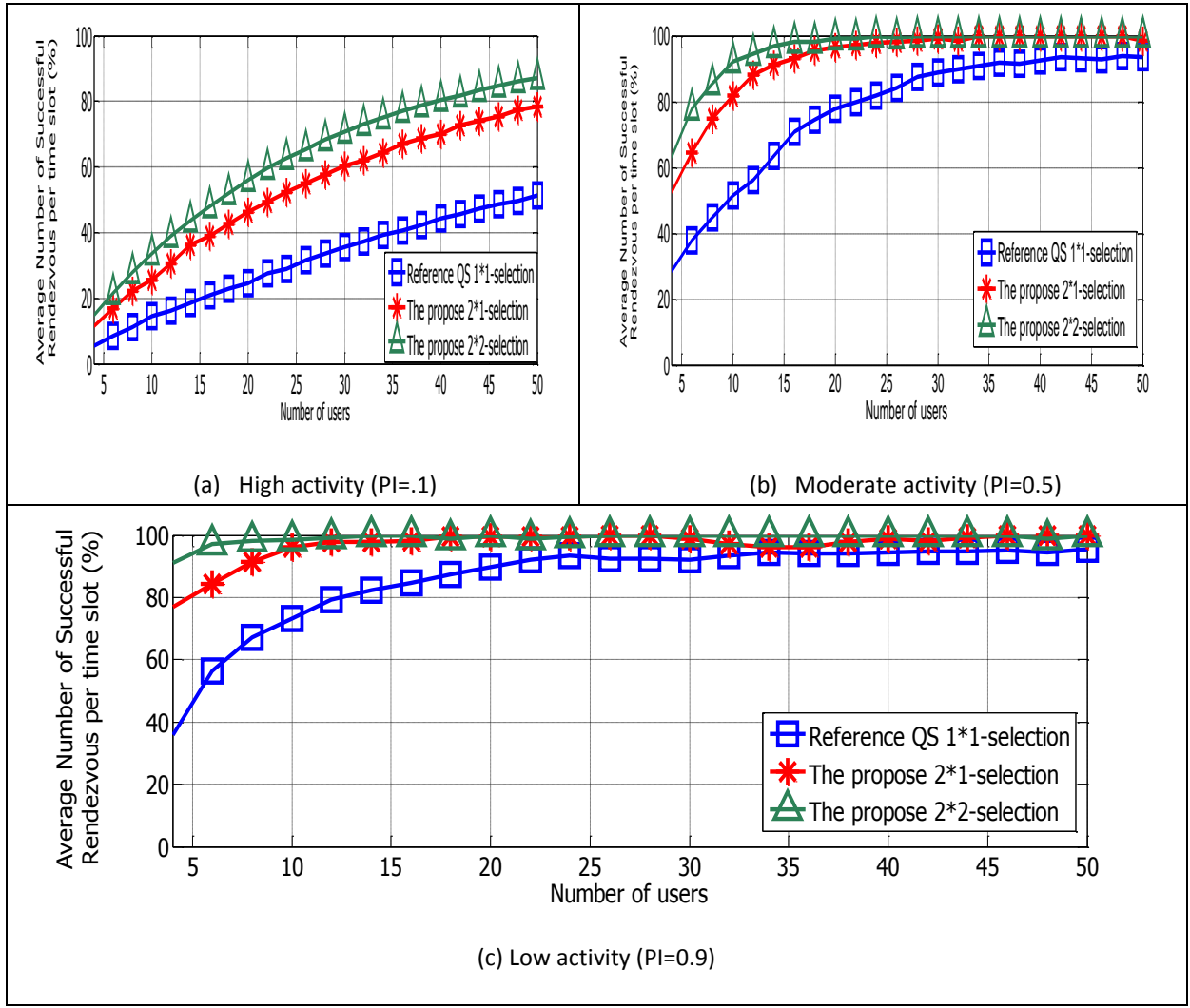

Fig.41: Average number of successful rendezvous per time slot vs. number of users, for grid size=3×3.

For all grid sizes, we can observe that the selection of 2-rows and 2-columns achieves the best performance. This is expected as its quorum interval is longer than that of the others schemes, and hence the number of intersections between nodes increase. Also, the average number of rendezvous increases as the channel availability increases. In this case, the selection of one row and one column out performs the 2×1 and 2×2-seletion schemes as expected because the intersection between any two CH sequences is small. Hence, it is more susceptible against rendezvous failures, which is caused by PU appearance.

Figure 41 shows the average number of rendezvous for grid size 3×3 and 9 available channels. It is obvious that the selection of two rows and two columns outperforms the others schemes for different $P_I$. In Figures 41(a), we note that the maximum rendezvous that can be achieved for 50 SUs is between 51% to 87%.

Figure 41 (b) and (c) show that all schemes achieve better average number of rendezvous as the number of SUs increases because the idle channel state becomes longer.

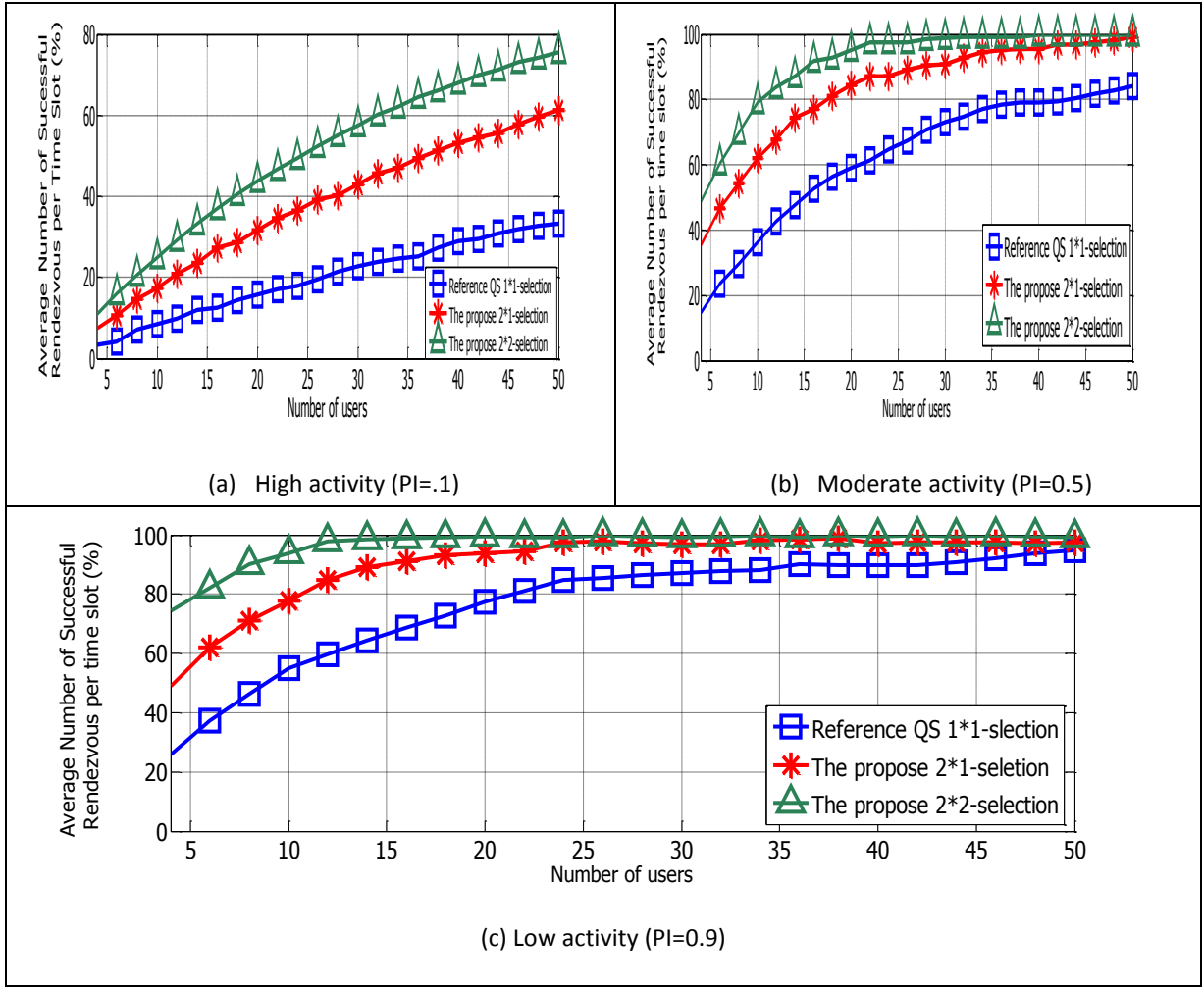

Fig.42: Average number of successful rendezvous per time slot vs. number of users, for grid size=4×4.

Figure 42 shows the average number of rendezvous for grid size 4×4 and 16 channels. Figure 42(a) shows that the 2×2-selection outperforms the other schemes under high activity when the number of SUs=50, the average number of rendezvous is around 33% to 76%). Figures 42(b) and (c) also show that the 2×2-selection outperforms the other schemes under moderate and low PR activities. We can also observe that when selecting more rows and columns and PI increases, the values of rendezvous become comparable. For $P_I$=0.9, we can observe that the 2×2 and 2×1-selections have comparable average number of successful meetings as the 2×2-selection enables almost 100% of node pairs to rendezvous when SUs ≥ 20, where the average RDV is ≥ 95% for 2×1-selection when SUs ≥ 30 as observed in Figure 42(c).

Next, we study the normalized average number of successful RDVs per quorum as a function of $P_I$ a shown in Figures 43-45. Figure 43 shows the performance when the grid size=3×3 and 9 channels for different number of SUs. It is obvious that the 2×2-seletion outperforms the other two selections since the channel availability increases, and hence the normalized average number of successful RDVs per quorum increase.

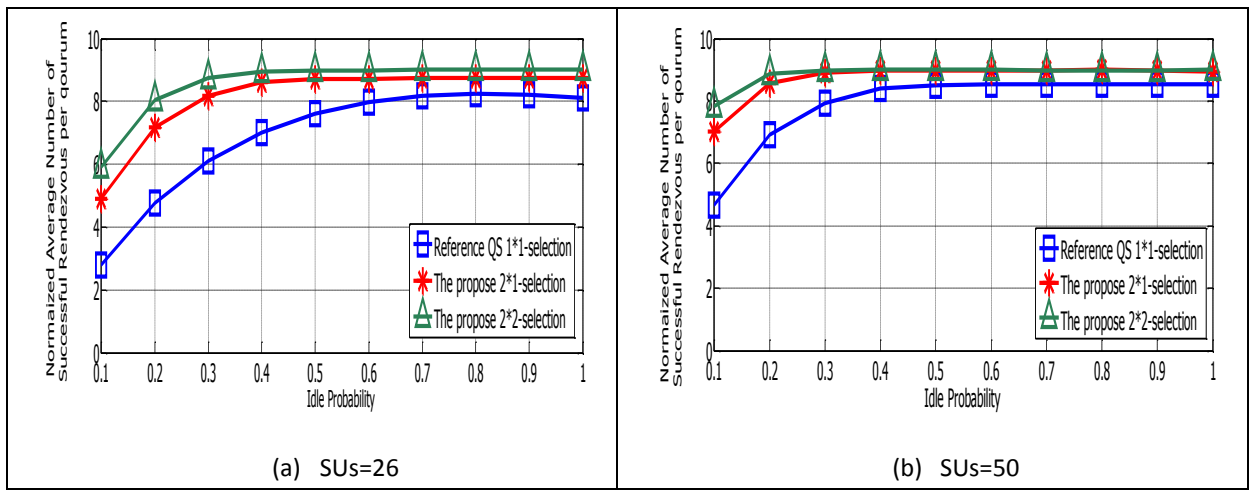

(a) SUs=26    (b) SUs=50

Fig.43: Normalized average number of successful rendezvous per quorum vs. $P_I$, for grid size=3×3.

Figures 44 and 45 show the normalized average number of successful RDVs per quorum for grid size=4×4 and 5×5. It can be noticed that as the grid size increases, the normalized average number of successful RDVs per quorum increases. The 2×2 and 2×1-seletions have comparable results when $P_I$ and number of SUs increase.

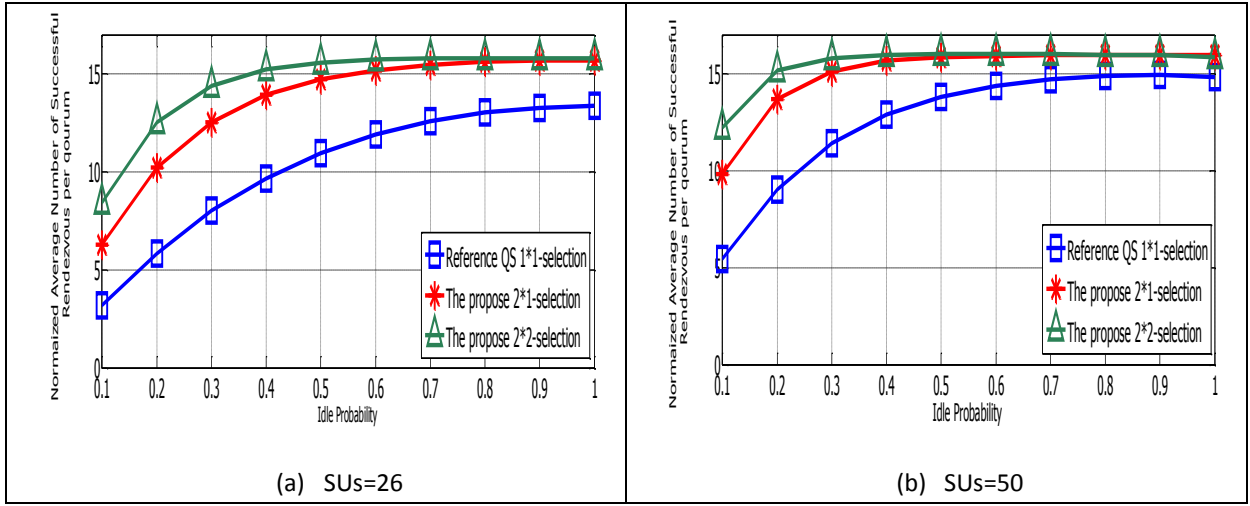

Fig.44: Normalized average number of successful rendezvous per quorum vs. P$_I$, for grid size=4×4.

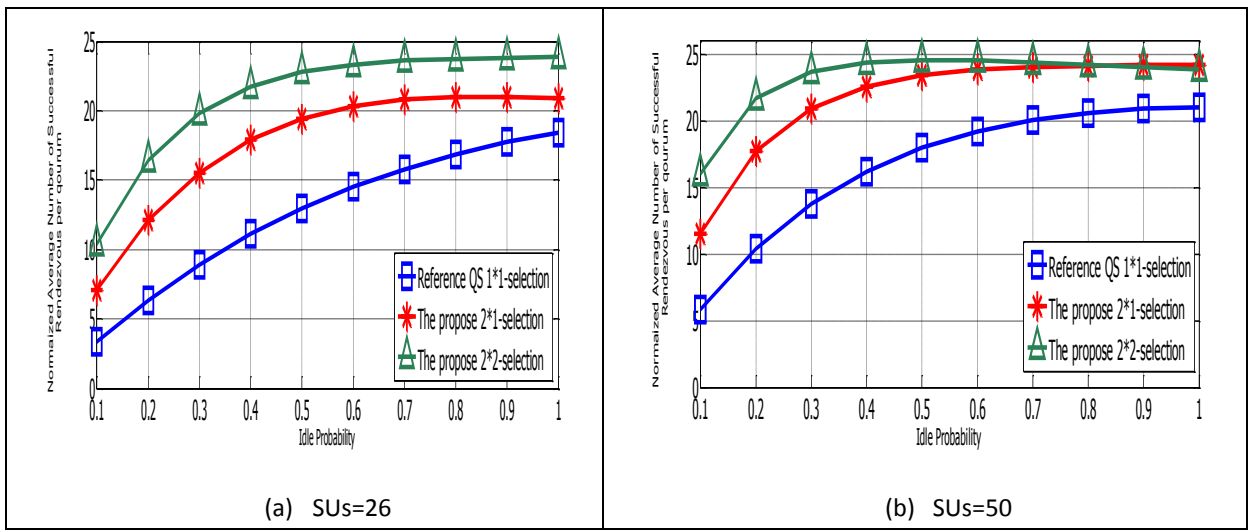

Fig.45: Normalized average number of successful rendezvous per quorum vs. P$_I$, for grid size=5×5.

Next, we study the normalized average number of successful rendezvous per quorum as a function of the number of SUs in Figures 46 and 47 for different PRs activities.

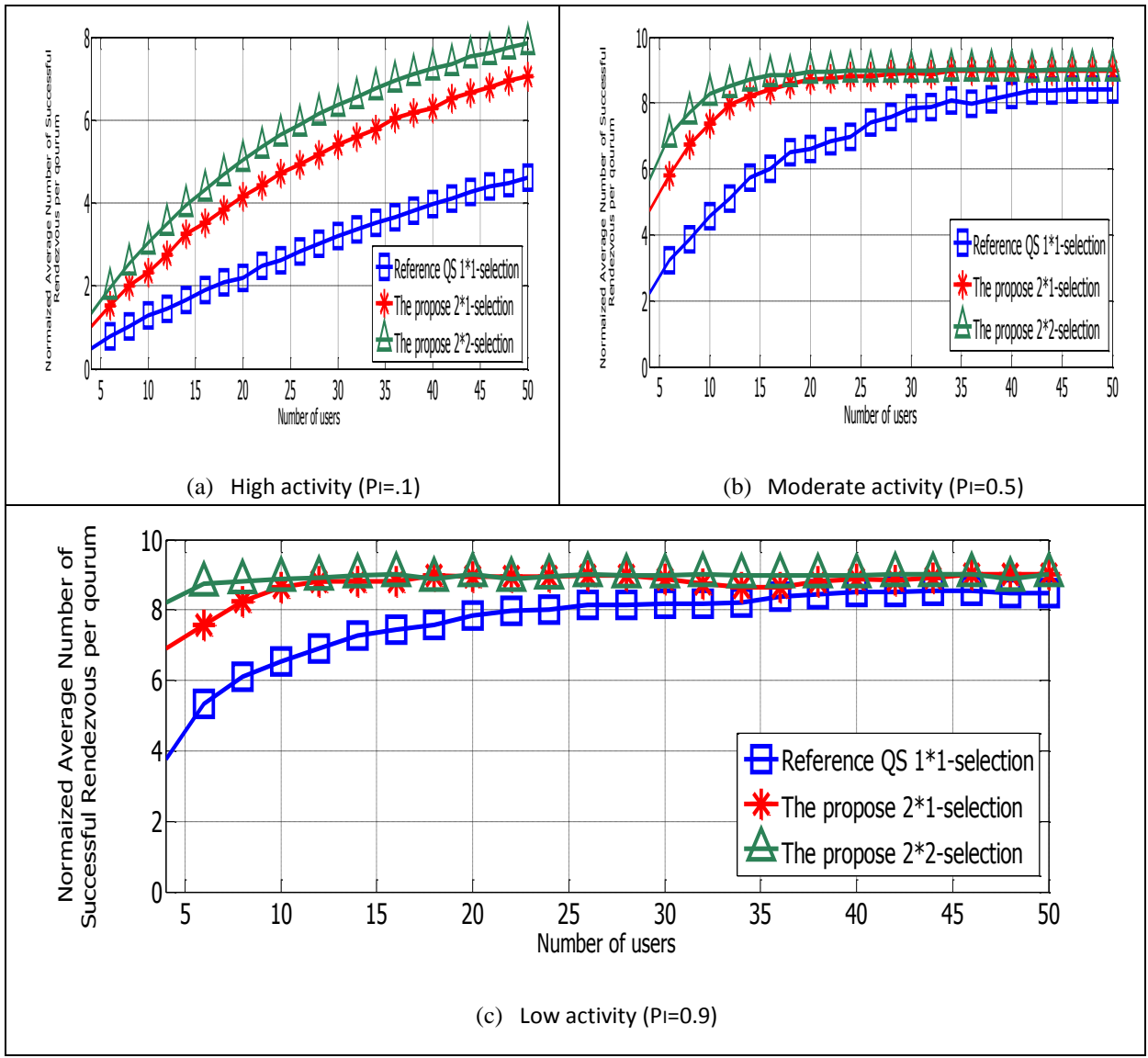

Fig.46: Normalized average number of successful rendezvous per quorum vs. number of users , for grid size=3×3.

Figures 46(a) and (c) show the results when the grid size=3×3 and the number of channels are 9. If when the three selections are compared for different channel availability, we find that the 2×2-selection outperforms the other two selections. As the number of SUs increases, the 2×1-selection almost achieves the same performance as the 2×2-selection.

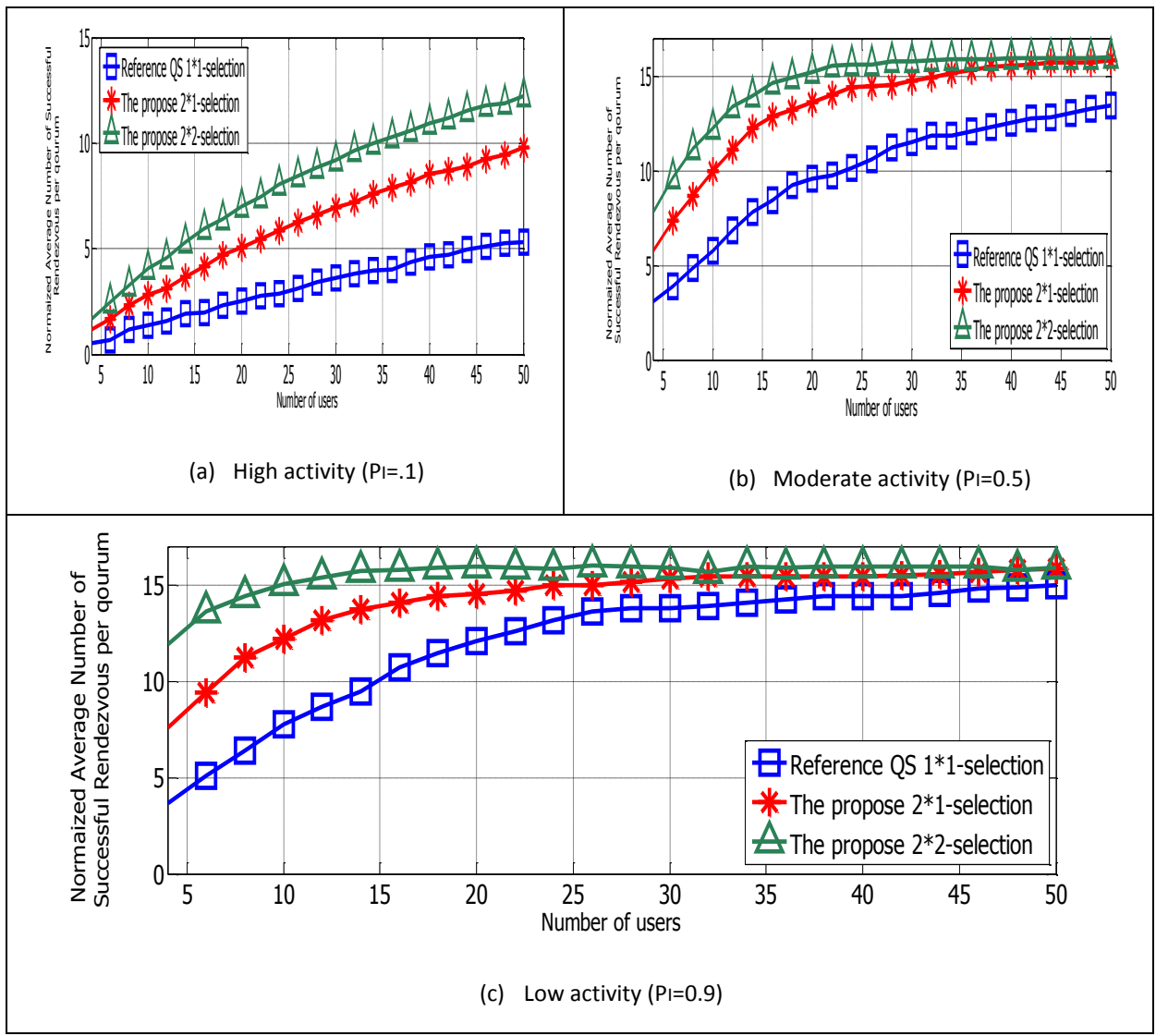

Fig.47: Normalized average number of successful rendezvous per quorum vs. Number of users, for grid size=4×4.

From Figures 47(a-c), it can be observed that 2×2-selection improve the performance per quorum. Under low PR activity at $P_I$=0.9, Figure 47(c) shows that when (SUs≥24) 2×1-selection has almost similar performance as 2×2-selection.

## 3.2.2 Average TTR

Next, we study the average TTR of SU pairs to rendezvous. When the average number of successful RDVs decreases (or the PRs activity increases), the average TTR will rise due to the less availability of common channels between SU pairs. Figures 48-53 show the average TTR comparison for the different schemes. It is obvious that when increasing the selection of the number of rows and columns, the quorum interval for each SU becomes longer so more intersections occur, which results in a lower average TTR.

The average TTR is investigated as a function of $P_I$ for different grid size and number of SUs in Figures 48-51. For all schemes, we observe that as the $P_I$ increases, the available channels increase, which leads to lower average TTRs (the 2×2-selection achieves the lowest average TTR compared to the other schemes). In this thesis, each time slot is enough to exchange RTS (160 bits), CTS (112 bits) and DTS (112 bits) control messages, where the time slot duration for successful exchange is 0.384 ms. The average TTR in ms for grid quorum system with size 3×3 and 9 available channels is investigated as in Figure 48. From now, we use the term of time slot to calculate the average TTR.

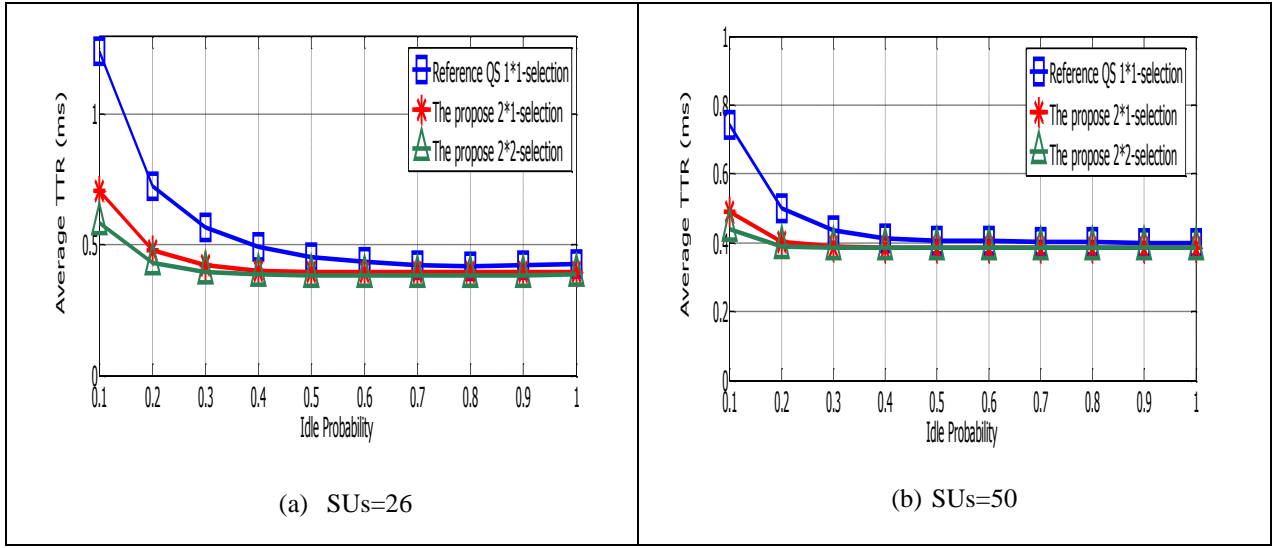

Fig.48: Average TTR (ms) vs. $P_I$, for grid size=3×3.

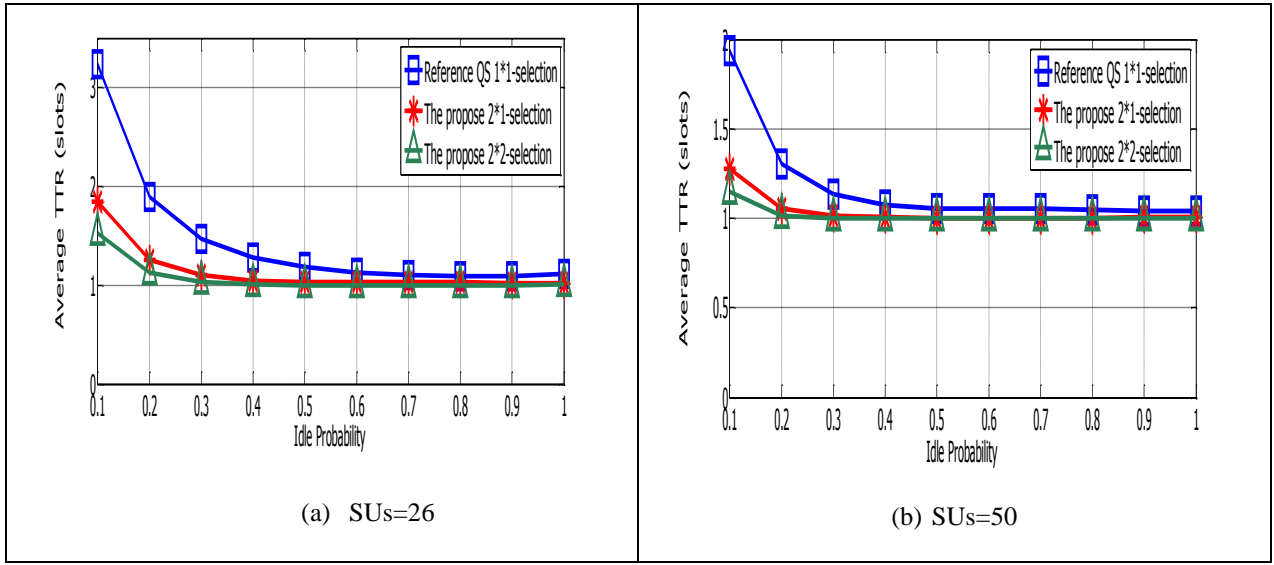

Fig.49: Average TTR (slots) vs. $P_I$, for grid size=3×3.

First, the grid quorum system with size 3×3 and 9 channels is investigated in Figure 49. When the 1×1-selection is compared to the 2×1 and 2×2-selections, we observe that the 1×1-selcetion introduces more average TTR than the others. This is expected due to the increase in the quorum interval and the larger availability of channels (i.e., more intersections between each pair of SUs). The average TTR for SUs=26 is almost comparable to that when the number of SUs=50 for the 1×1-selection. Also, the 2×1-selection achieves similar performance as the 2×2-selection when $P_I>0.6$.

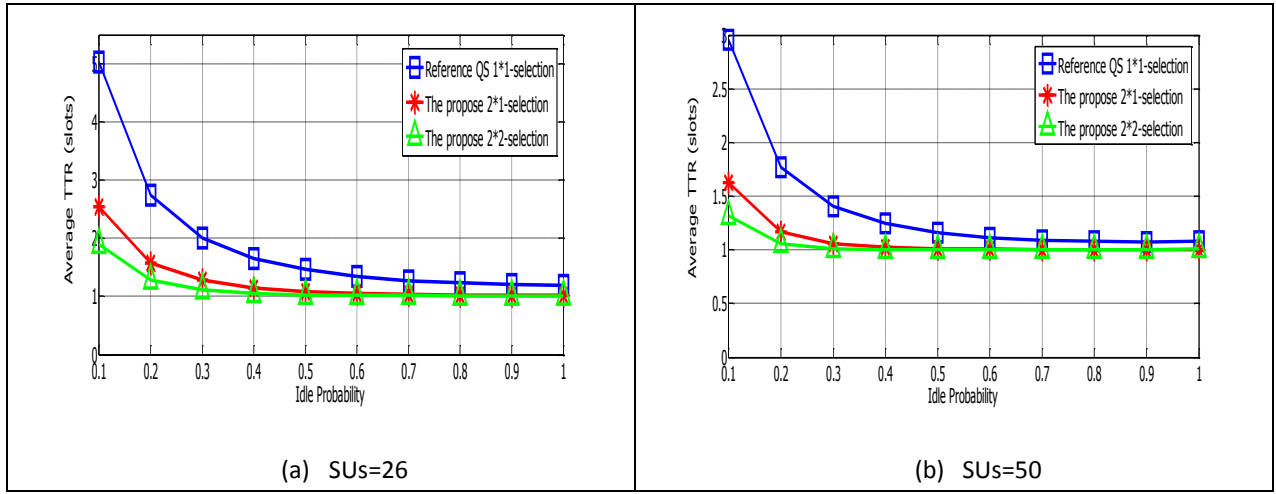

(a) SUs=26

(b) SUs=50

Fig.50: Average TTR (slots) vs. $P_I$, for grid size=4×4.

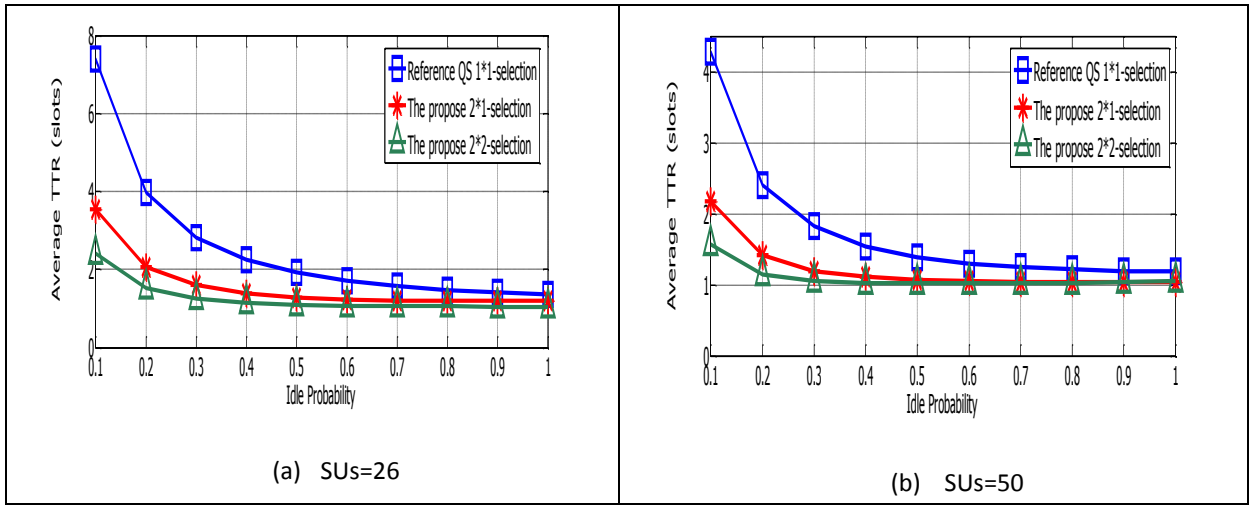

(a) SUs=26

(b) SUs=50

Fig.51: Average TTR (slots) vs. $P_I$, for grid size=5×5.

We now consider the average TTR as a function of $P_I$ for grid size = 4×4 and 5×5, respectively. Figures 50 and 51 reveal that similar behavior is observed for the two grid size, but with higher average TTR.

Figures 52 and 53 show the average TTR as a function of the number of SUs under different values of $P_I$. When the average number of successful RDV decreases, the average TTR will increase due to the less common available channels between the SU.

Now, we study the grid quorum system with size=3×3 and 9 channels in Figure 52. The average TTR decreases as $P_I$ increases due to the fact that when the availability of the channel is high, SUs pairs achieve RDV without encountering PU effects. Note that the 2×2-selection achieves the best average TTR performance among other two scheme.

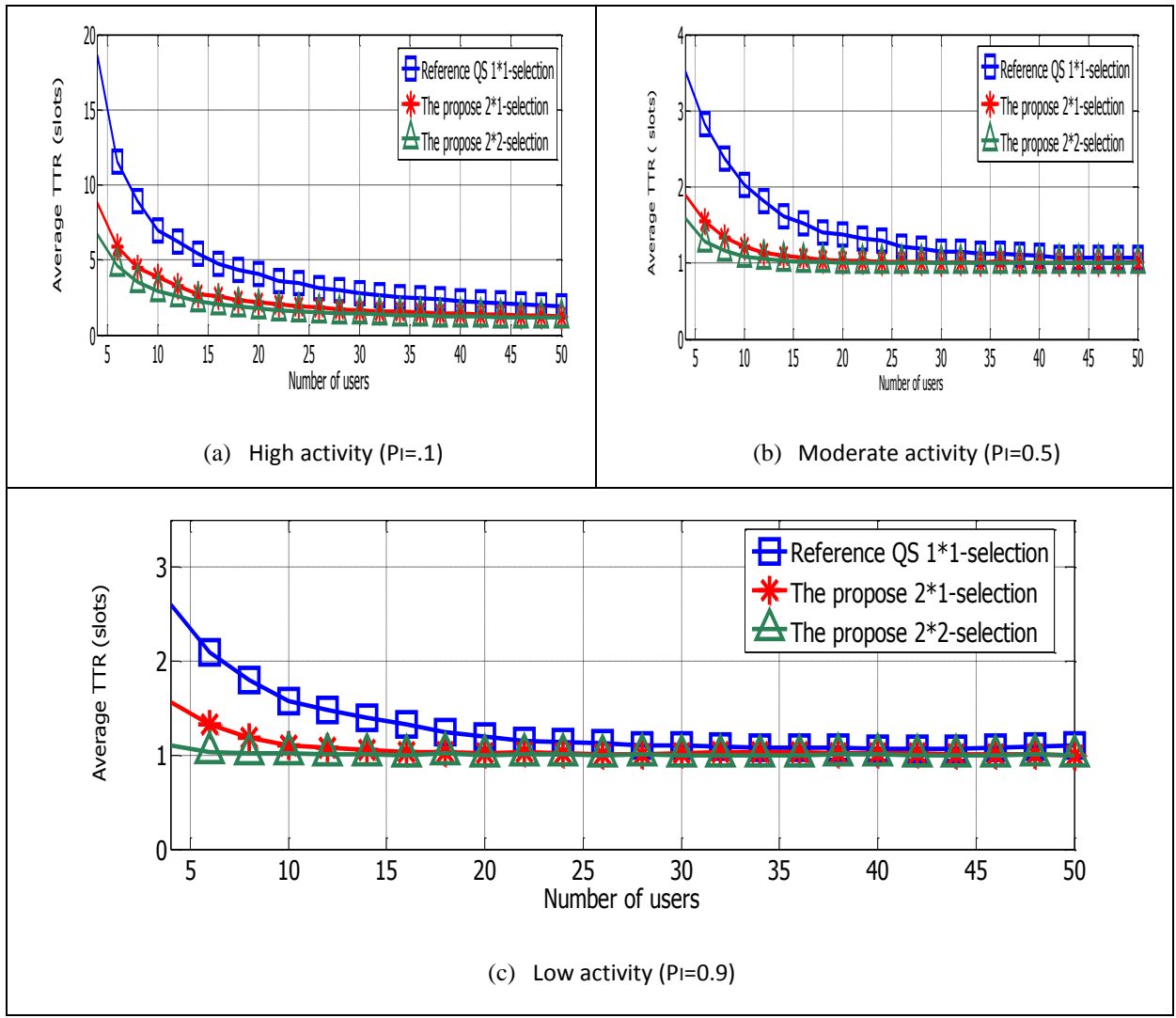

Fig.52: Average TTR (slots) vs. number of users, for grid size=3×3.

Next, Figure 53 shows the average TTR for different channel availability with grid size= 4×4. We observe that as the grid size increases, the average TTR increases due to the fact that the CH sequence increases and so SUs pairs take more time slots to RDV. As expected, the 2×2-selection has the best performance under the different channel availability due to the fact that it has more intersection slots.

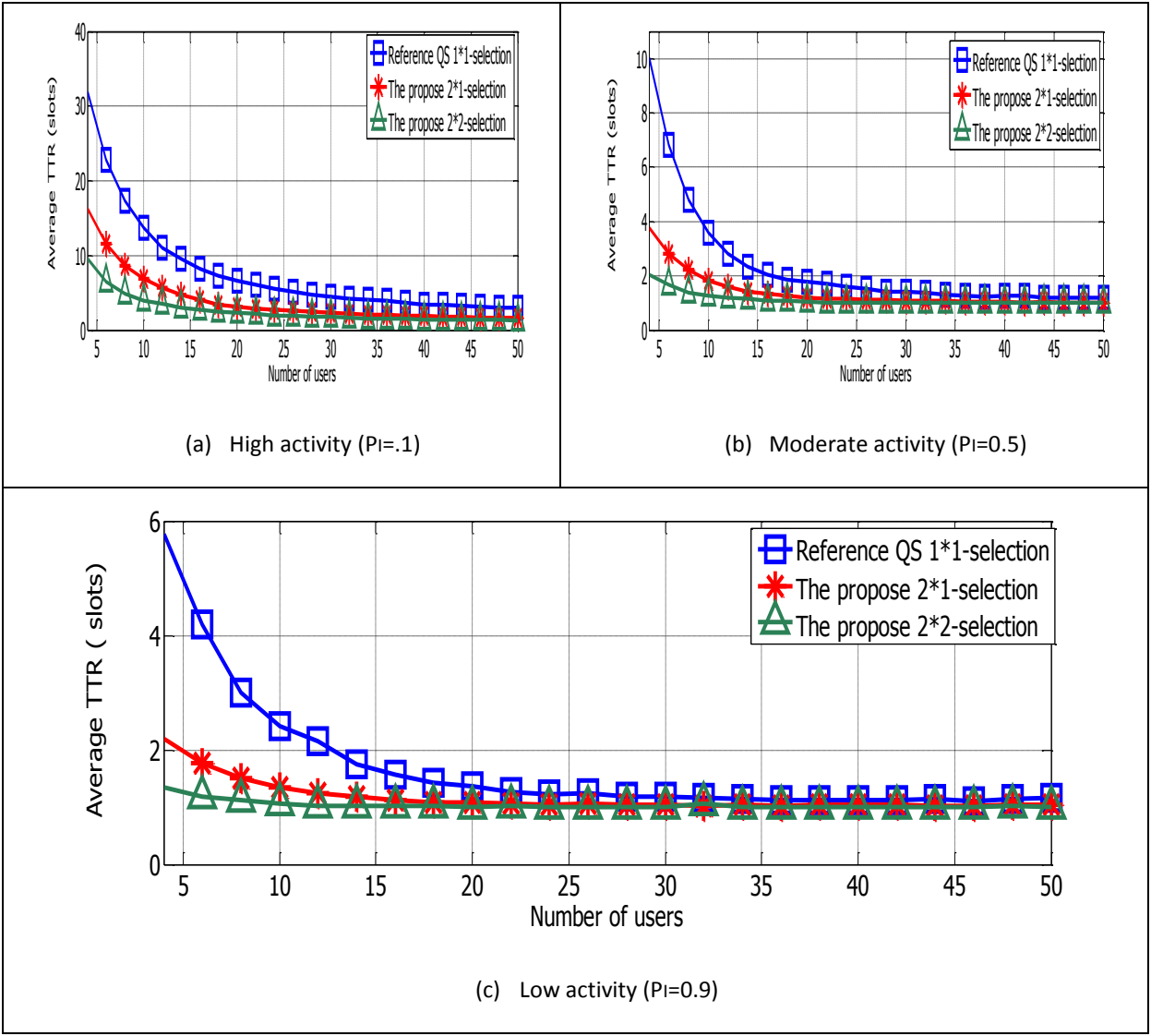

Fig.53: Average TTR (slots) vs. number of users, for grid size=4×4.

### 3.2.3 Normalized Energy per Successful RDV

Next, we study the normalized energy consumption per a successful RDV, which indicates the number of units of energy consumed per a successful exchange assuming that each time slot consumes 1 unit of energy:

$$\text{Normalized energy per a successful RDV} = \frac{Total\ number\ of\ active\ slots}{Total\ number\ of\ RDV} \times 1\ unit\ of\ energy$$

Figures 53-55 show the normalized energy per successful RDV as a function of $P_I$ for grid size = 3×3, 4×4 and 5×5, respectively. The 2×2-seletion has the longest quorum interval, the number of active slots will be greater compared to the other schemes. From Figure 53, we observe that for SUs=26 as long as ($P_I$<0.2), the 1×1-selection consumes more energy per successful transmission. This can be attributed to the fact that at low channel availability the 1×1-selection has minimum successful RDVs compared to the other schemes. For $P_I$>0.2, the increase of the number of rows and columns will increase the amount of needed energy per a successful transmission where the 2×2-selection consumes more energy (the 1×1-seletion consumes lower amount of energy). We also notice that almost the same amount of energy is consumed when $P_I$>0.2 for the 2×2-selection, $P_I$>0.3 for the 2×1-selection and when $P_I$>0.5 for the 1×1-selection. For SUs=50 at high activity of PUs, the three schemes consume a comparable amount of energy and as ($P_I$>0.1), the 2×2-selection also consumes more energy than the others schemes. Compared to Figure 53 (b), the three schemes almost consume the same amount of energy at small $P_I$, where $P_I$>0.2 for the 2×2-selection, $P_I$>0.2 for the 2×1-selection and when $P_I$>0.4. This is expected due to the increase in the number of SUs.

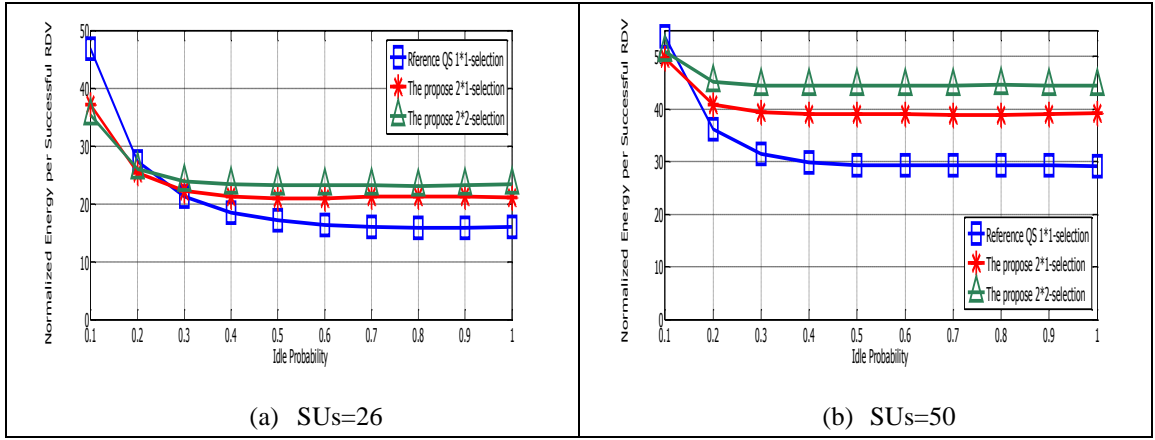

(a) SUs=26        (b) SUs=50

Fig.53: Normalized energy per successful RDV vs. $P_I$, for grid size=3×3.

Next, we consider the normalized energy per a successful RDV for grid size=4×4. We observe that the 1×1- selection consumes more units of energy when $P_I$<0.3, and decreases with $P_I$ up to $P_I$=0.7, where the consumed energy will be almost the same. The 2×1-selection consumes more energy than the 2×2-selection for $P_I$>0.2 and also it decreases until $P_I$=0.5, where it consumes the same amount of energy. Figure 54 (b) reveals the same performance as the one in Figure 54 (a), but with higher energy consumption for all schemes.

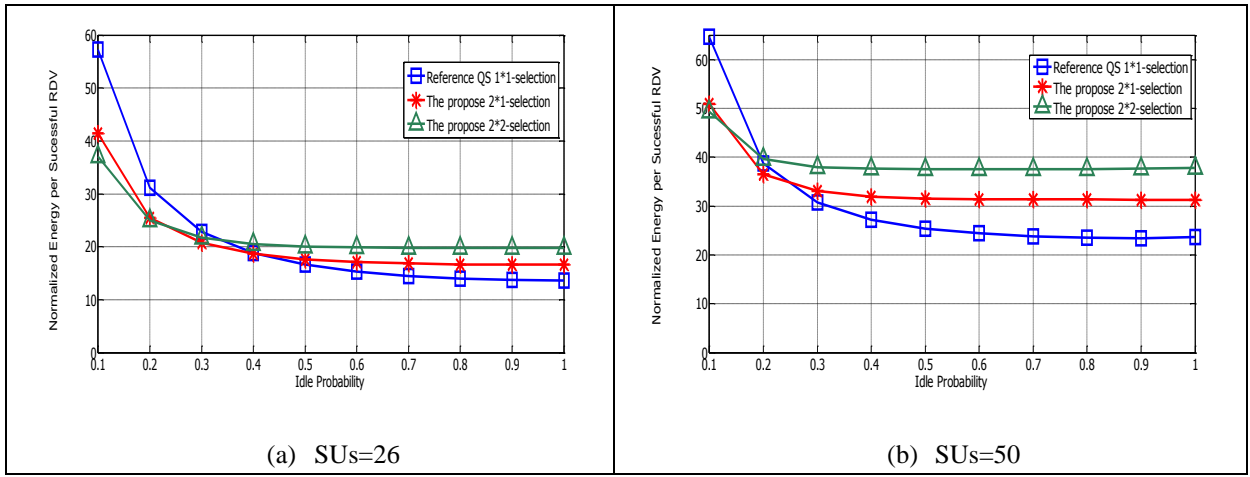

(a) SUs=26　　　(b) SUs=50

Fig.54: Normalized energy per successful RDV vs. P$_I$ for grid size=4×4.

Figure 55 shows that normalized energy consumption per a successful RDV for grid size= 5×5. The consumed energy consumed per a successful RDV in this case for higher PR activities are greater than when the grid size= 3×3 and 4×4. This is because as the grid size increases, the number of active slot increases.

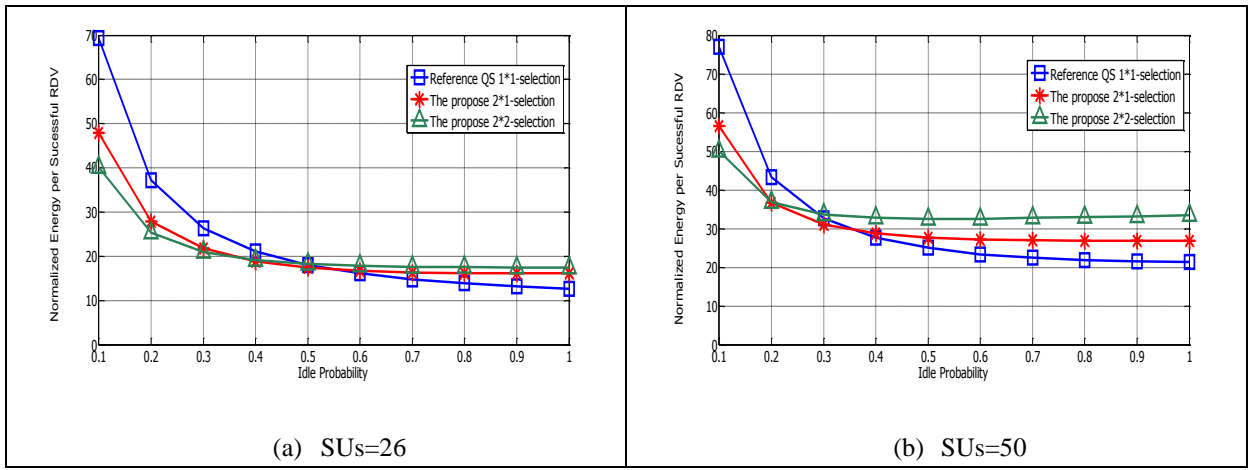

(a) SUs=26　　　(b) SUs=50

Fig.55: Normalized energy per successful RDV vs. P$_I$ for grid size=5×5.

Figures 56 and 57 show the normalized energy per a successful RDV as a function of number of SUs. Figure 56 (a) shows that the 1×1-selection consumes more energy than the other selections. We also notice that the energy consumption increases linearly with the number of SUs. The 2×1 and 2×2-selection consume comparable amount of energy. For $P_I$=0.5, we can observe that as long as the number of SUs is less than 20, the three schemes achieve comparable performance. As the number of SUs increases, the 2×2-selection consumes more energy.

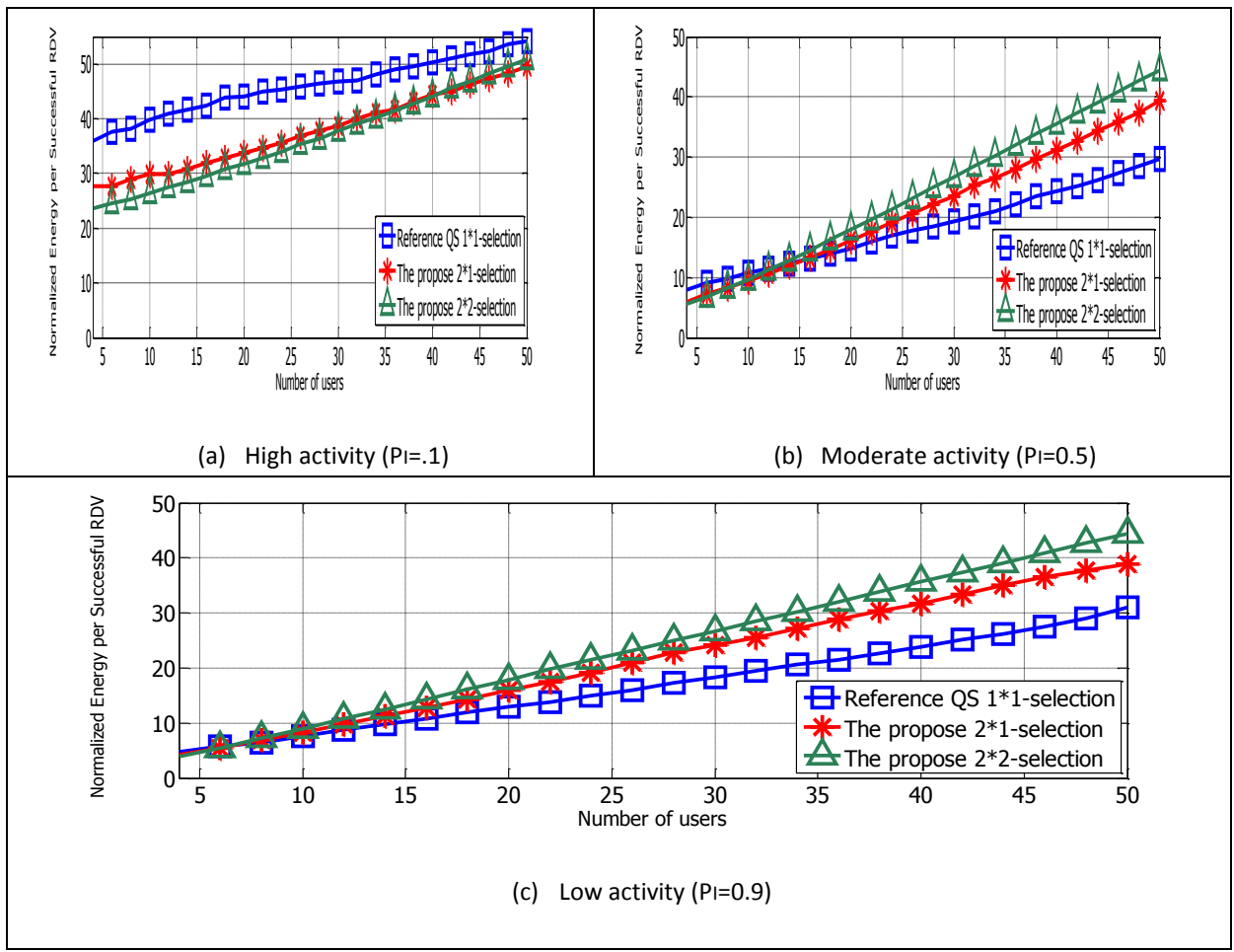

Fig.56: Normalized energy per successful RDV vs. number of users, for grid size=3×3.

Figure 57(a) shows that the 1×1-selection consumes more energy because of the number of RDVs is almost the same. We can notice that at ($P_I$=0.5), the 3×3 and 4×4 have comparable performance when SUs<20. As the number of users increases, the 4×4 consumes less energy per a successful RDV. This is expected due to the fact that as the grid size increases, the chance of serving more users increases and hence the number of RDVs increases. This reduces energy consumption per a successful RDV. Figure 57(c) indicates the performance when $P_I=0.9$, which is similar to the performance when $P_I=0.5$ (comparable performance for all schemes when SUs< 15).

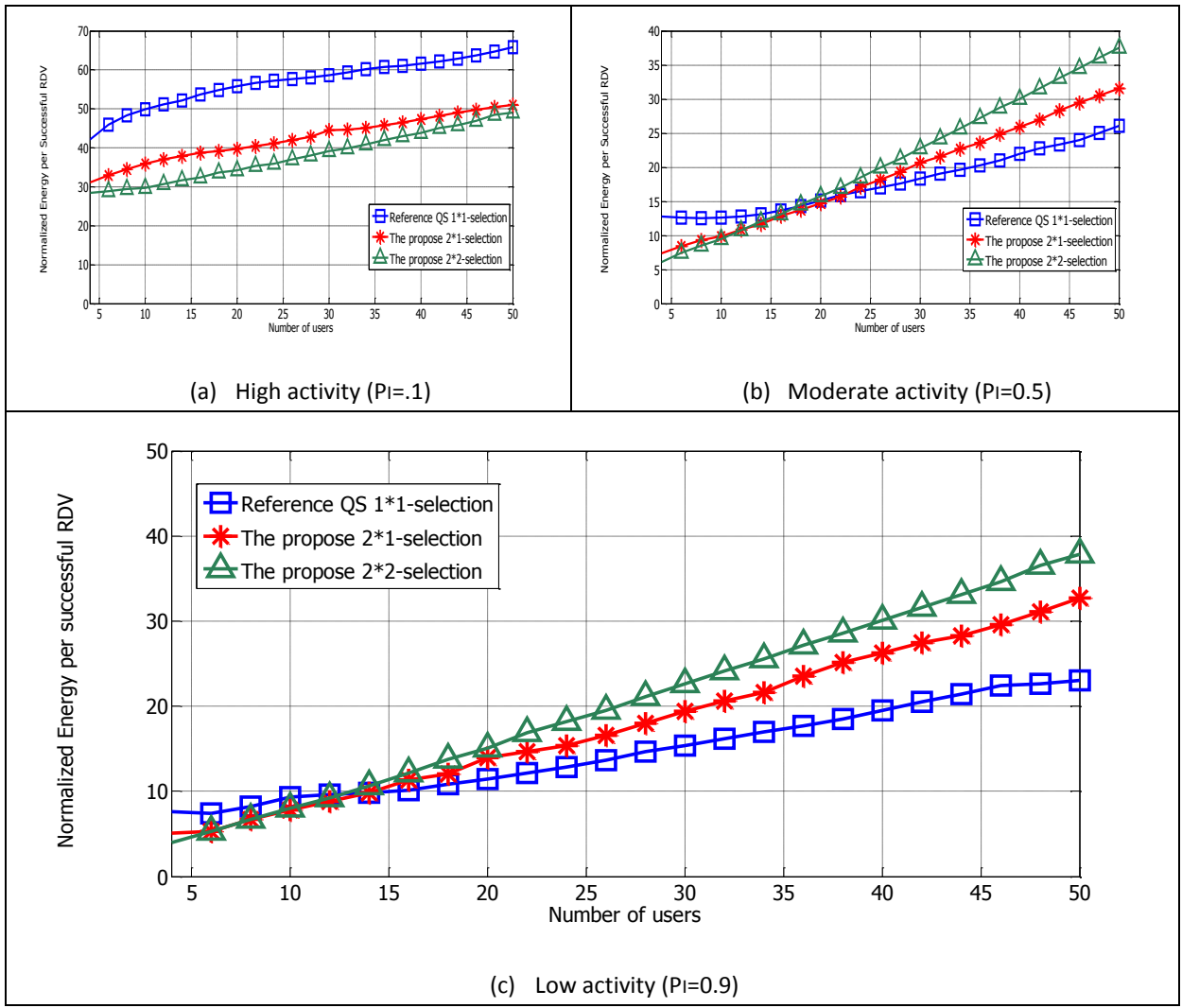

Fig.57: Normalized energy per successful RDV vs. number of users, for grid size=4×4.

## 3.2.4 Forced Blocking:

The last performance metric is the forced blocking probability, which is the probability to forced blocked SUs due to the PU activity. The forced blocking probability of SUs can be expressed as the average number of blocking per time slot of SUs divided by the average number of successful meetings per time slot:

$$\text{Forced Blocking Probability} = \frac{Average\ number\ of\ blocking}{Average\ number\ of\ successful\ meeting}$$

Figures 57-59 show that similar behavior is observed for all schemes for different grid sizes and different number of SUs. This is expected and can be observed in the Figures whereas. Note that as $P_I$ increases, the forced blocking probability decreases dramatically. This is because at low $P_I$ the number of channels will not be sufficient to support any SU transmission and so new SUs will be blocked. At high $P_I$, the number of available channels increases, so the communications between SUs will also increase. Thus, we can conclude that the forced blocking depends on the availability of the channels and PR activity, but it does not depend on the grid size.

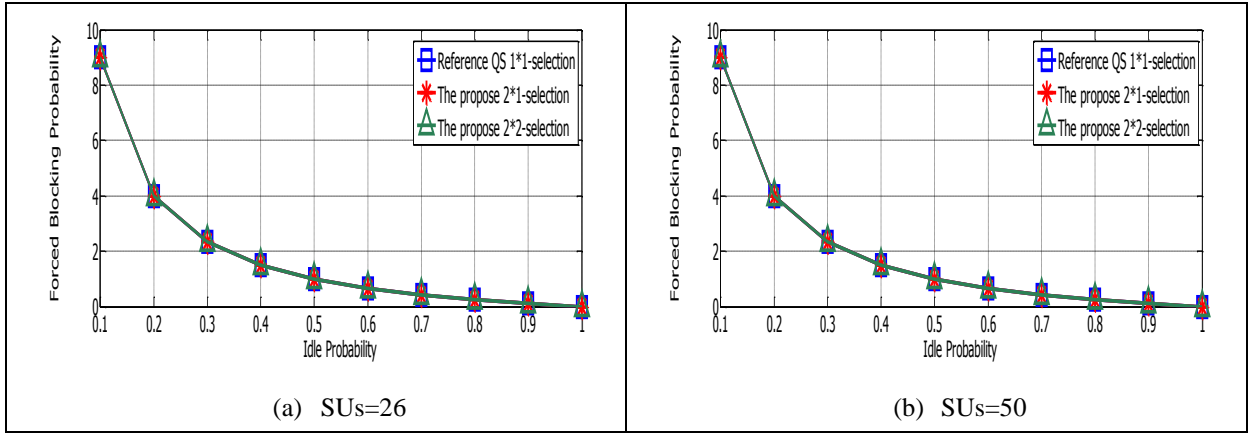

(a) SUs=26      (b) SUs=50

Fig.58: Forced blocking vs. $P_I$, for grid size=3×3.

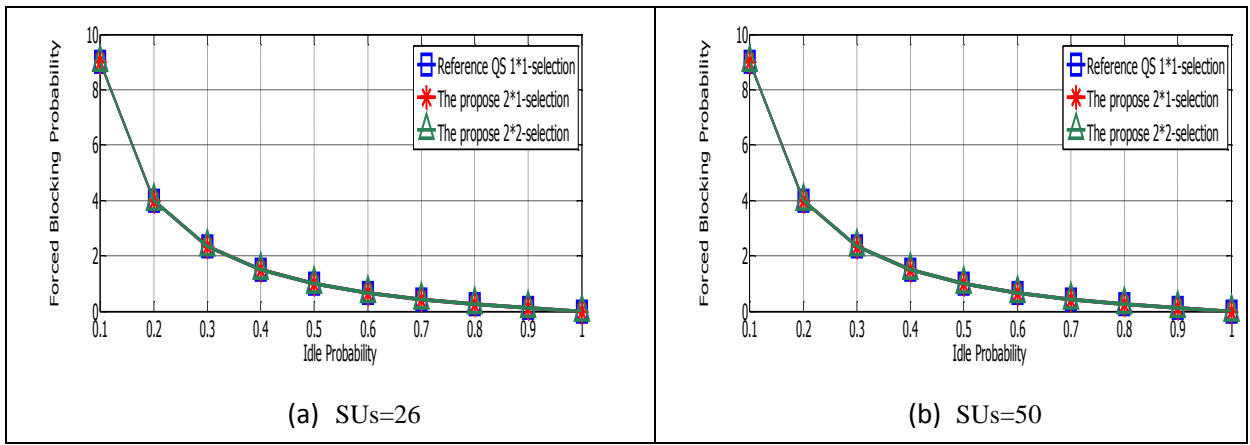

(a) SUs=26      (b) SUs=50

Fig.59: Forced blocking vs. $P_I$, for grid size=4×4.

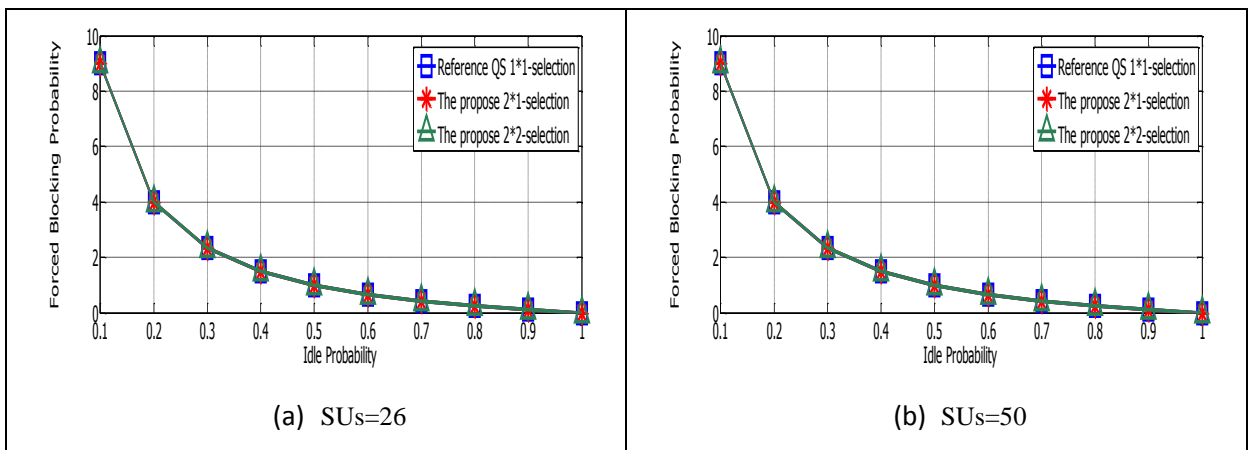

(a) SUs=26      (b) SUs=50

Fig.60: Forced blocking vs. $P_I$, for grid size=5×5.

## 3.2.5 Performance Evaluation of Adaptive Algorithm:

We need an adaptive approach to improve the performance in terms of number of RDV, average TTR and the number of units energy consumed per successful RDV, so it's necessary to allow different CR users to dynamically select appropriate rows and columns from the quorum system based on the traffic load and network performance requirements. From the previous results we observed that the 2×2-selection always gives the best performance of number of successful RDV but consumed more energy, 1×1-selection consumed less energy and achieved less number of successful RDV and 2×1-selection consumed less energy than the 2×2-selection and achieved more number of successful RDV than the 1×1-selection.

For each metric we evaluated the adaptive algorithm for different number of users to achieve best performance of average number of successful RDV with less energy consumption and study the effect of varying the availability on channel on different number of users for grid size 4×4.

Figures 61 shows the average number of rendezvous performance, we can observed that our adaptive algorithm achieved performance approximate to the 2×2-selection.

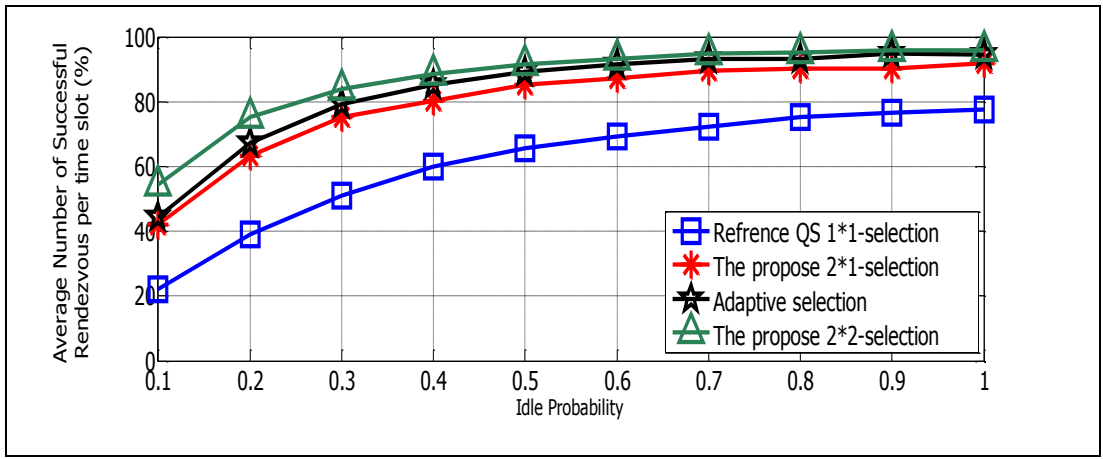

Fig.61: Performance evaluation: average number of successful rendezvous per time slot vs. $P_I$, for grid size=4×4.

Next, we study average TTR metrics we observe that our adaptive algorithm gives performance between 2×1-selection and 2×2-selection when $P_I$ <0.4 and after that the adaptive algorithm achieves similar performance as 2×1-selection and 2×2-selection as shown in Figure 62.

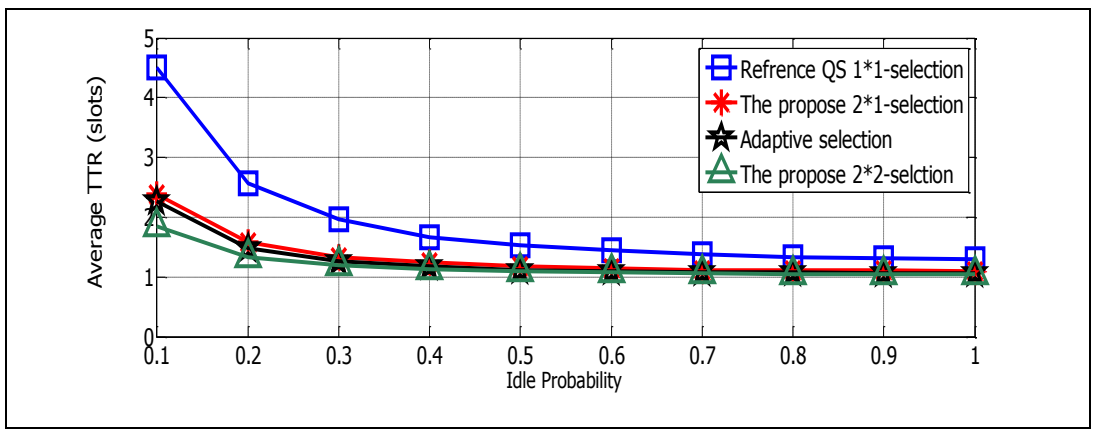

Fig.62: Performance evaluation: average TTR (slots) vs. $P_I$, for grid size=4×4.

Figure 63 shows the normalized energy per successful RDV, we can noticed that the adaptive algorithm get similar performance as 2×1-selection.

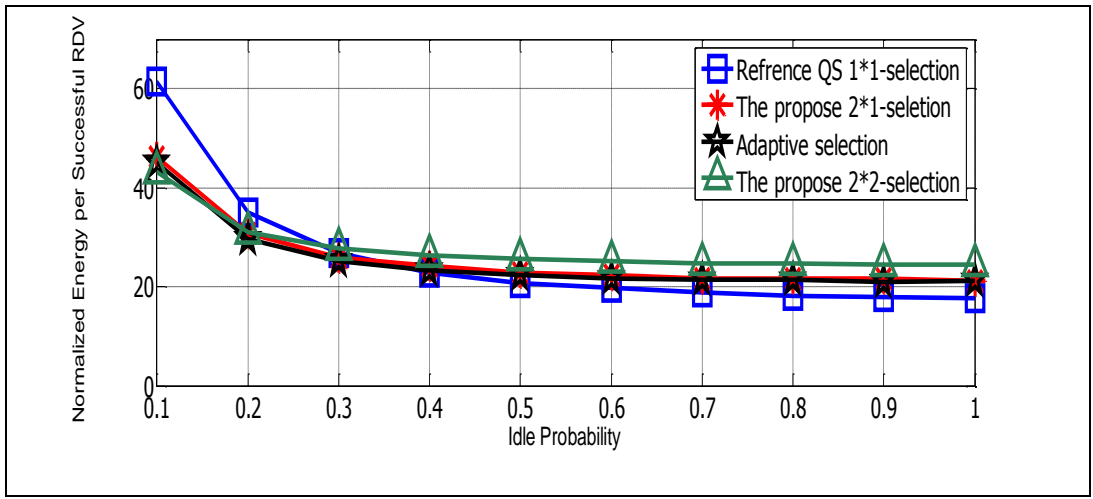

Fig.63: Performance evaluation: normalized energy per successful RDV vs. P$_I$, for grid size=4×4.

Figure 64 shows the forced blocking probability, we observe that the adaptive algorithm has the similar behavior to all scheme which is expected as mention before.

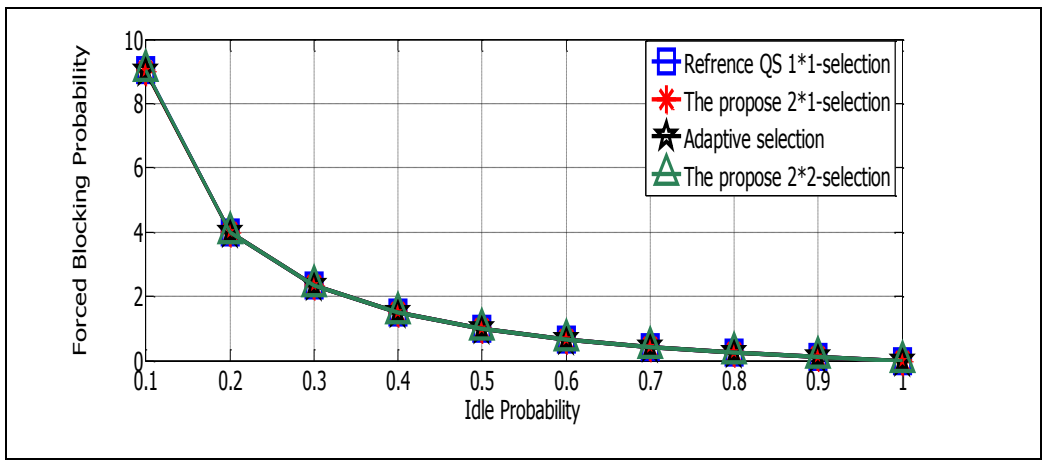

Fig.64: Performance evaluation: forced blocking vs. P$_I$, for grid size=4×4.

# Chapter 4: Conclusion and Future Work:

## 4.1 Conclusion:

In this thesis, we have studied the grid quorum system to develop new channel-hopping-based distributed rendezvous algorithms. We proposed and analyzed four different channel selection algorithms three of them based on selecting variable number of rows and columns: (1) one-row and one-column selection, (2) two-rows and one-column selection or one-row and two-columns selection (3) two-rows and two-columns selection and (4) adaptive number of rows and columns selection. The results of these selections are used by each SU to generate its own channel hopping sequence. We solve the RDV problem by increasing the number of common channels between any pair of CH sequence.

We evaluated the performance of our proposed algorithms through extensive simulation experiments using Matlab. The performance analysis verified that there are tradeoff between the four metrics for different selection schemes (1×1, 2×1 and 2×2). The performance analysis showed that the most important factor that effects the performance is the PR activity. When the PRs activity is higher, the average number of RDVs decreases, the average TTR increases and the forced blocking increases. Simulation results also showed that as the number of rows and columns increases most of the RDVs occur and lower average TTR is achieved, in which the 2×2 achieves the best performance but consumed more energy per a successful RDV for high channel availability. If saving energy is our main objective, with almost the same amount of average number of RDVs (as the 2×2-algorithm), we can use the 2×1-algorithm. The performance analysis showed that at low channel availability the 1×1-algorithm consumed more energy per a successful

RDV whereas the 2×1 and 2×2-algorithms consumed almost the same amount of energy per a successful RDV. In addition, our results also showed that as the grid size increases as a function of $P_I$, the average number of RDVs decreases and less amount of energy per a successful RDV is consumed. The performance analysis also showed that the forced blocking depends on the availability of the channels and the PR activities, but it does not depend on the grid size.

In summary, we need an adaptive approach, which dynamically adapts the selection of the number of rows and columns to achieve the best required performance for our network based on the main objective (i.e., the average number of RDVs or the amount of energy consumption per a successful RDV.

## 4.2 Future Work:

Our channel selection algorithms can be extended to accommodate the scenario, in which the SUs experience asymmetric channel views. This work can be investigated to estimate the channel quality, which provides instantaneous and historical measurements. In addition, we plan to extend our work to the multi-hop scenario, where the problem is more challenging due to the synchronization requirements. One possibility of adaptation that have quality of service requirement in terms of average TTR, average number of successful RDV or energy consumption. This work can be extend to select the threshold of traffic load dynamically.

# الملخص العربي

**العنوان:** لأغراض شبكة راديوية ذكية على اساس نصاب قانوني مكيف لتنقل القناة و ذلك بنظام تنسيق موزع
**الاسم:** إسراء زياد نواف الجراح (2013976006).
**المشرف:** د. هيثم بني سلامه.
**المشرف المشارك:** د. علي اعيدة

من اهم التحديات في انتشار الشبكات الراديوية الإدراكية (CRNs) هو ايجاد قناة تحكم مشتركة (CCC) لجميع المستخدمين الثانويين (SUs) التي تتيح الاتصال الفعال بين مستخدمين CR . يعزى هذا التحدي الى التغيير الديناميكي على فترات زمنية متفاوتة لظروف هيكل الشبكة و مواقع توفر الطيف. يعرف اللقاء على انه عملية انشاء اتصالات للتحكم والذي هو شرط اساسي لاي اتصال فعال بين اي نقطتين (CR). من اكثر بروتوكولات اللقاء بين مستخدمين CR انتشارا تستند الى (QSs). QSs يمتلك العديد من الخصائص الفعاله التي يمكن استخدامها لتأسيس اتصالات دون الحاجة لوجود قناة تحكم مشتركة و بالتالي التغلب على مشكلة اللقاء (RDV).

في هذا البحث عرضنا خوارزمية لقاء جديدة لتنقل القناة على اساس تقنيات شبكة QS. الخوارزمية المقترحة تزيد من احتمالية اللقاء (RDV) ضمن دورة واحدة عن طريق السماح لمستخدمين CR للقاء اكثر من مرة وفقا لخاصية التقاطع QS. تسمى الخوارزمية المقترحة " لأغراض شبكة راديوية ذكية على اساس نصاب قانوني مكيف لتنقل القناة و ذلك بنظام تنسيق موزع ". الفكرة الرئيسية للخوارزمية هو الضبط الديناميكي لاختيار QS من قبل مستخدمين CR وفقا للاعداد المتفاوته للمستخدمين في CRN. الخوارزمية المقترحة تقلل متوسط TTR و تزيد احتمالية اللقاء. قمنا بتقييم اداء الخوارزمية من خلال برنامج Matlab. تمت مقارنة الخوارزمية المقترحة مع نوعين مختلفين من مخططات QS. اظهرت النتائج ان الخوارزمية يمكن ان تقلل من TTR و تزيد من احتمالية RDV و تقلل من استهلاك الطاقة لكل RVD ناجح.